\documentclass[ twocolumn,aps,prd,   
               preprintnumbers,numbers,sort&compress,
               nofootinbib,
                            showpacs,
               colorlinks,
               linkcolor=blue,
               citecolor=blue]{revtex4-1}
   \newcommand{\exclude}[1]{}

\usepackage{graphicx,amsmath,amssymb,bm}
\usepackage{psfrag}
 \usepackage{feynmp}
\usepackage{hyperref}
\usepackage{enumitem}

\newcommand{\beq}{\begin{equation}}
\newcommand{\eeq}{\end{equation}}
\newcommand{\be}{\begin{eqnarray}}
\newcommand{\ee}{\end{eqnarray}}
   
\def\dd{ \,\mathrm{d} }

\def\+{\dagger}
 \def\la{\langle}
 \def\ra{\rangle}

\begin{document}
\title{\exclude{Aharonov-Bohm phases 
and  dynamical Casimir effect  
in a quantum LC circuit}Topological Casimir effect in a quantum LC circuit: real-time dynamics}

\author{Yuan Yao}
\affiliation{Department of Physics and Astronomy, University of British Columbia, Vancouver, B.C. V6T 1Z1, Canada} 
\author{Ariel R. Zhitnitsky}  
\affiliation{Department of Physics and Astronomy, University of British Columbia, Vancouver, B.C. V6T 1Z1, Canada}

\begin{abstract}
We study  novel contributions to  the  partition function  of  the Maxwell system  defined on a small compact manifold ${\mathbb{M}}$  with nontrivial mappings $\pi_1[U(1)]\cong\mathbb{Z}$. These contributions cannot be described in terms of conventional physical propagating photons with two transverse polarizations, and instead  emerge as a result of   tunneling transitions between topologically different but physically  identical vacuum winding states.\exclude{These new terms give an extra contribution to the Casimir pressure, yet to be measured.} We argue that if the same  system   is considered in the background of a small external time-dependent E\&M field,  then  real physical photons will be emitted from the vacuum, similar to the dynamical Casimir effect (DCE) where  photons are radiated  from the vacuum due to time-dependent  boundary conditions. The fundamental technical difficulty  for such an analysis is that the radiation of physical photons on mass shell is inherently a real-time Minkowskian phenomenon while the vacuum fluctuations interpolating between topological $|k\rangle$  sectors rest upon a Euclidean instanton formulation. We overcome this obstacle by introducing auxiliary topological fields which allows for a simple analytical continuation between Minkowski and  Euclidean  descriptions, and develop a quantum mechanical technique to compute these effects.
   
We also propose an experimental realization of such small effects using a microwave cavity with appropriate boundary conditions.  Finally, we comment on the possible cosmological implications of this effect.

 
\end{abstract}

\maketitle

\section{Introduction. Motivation.}\label{introduction}
 It has been recently argued \cite{Cao:2013na,Zhitnitsky:2013hba,Zhitnitsky:2014dra,Zhitnitsky:2015fpa,Cao:2015uza} that some novel  terms in the partition function emerge when pure Maxwell theory is defined on a small compact manifold.  These terms are not related to the  propagating photons with two transverse physical polarizations, which are responsible for the conventional Casimir effect (CE) \cite{Casimir}. Rather, they occur as a result  of  tunneling  events between topologically different but physically identical    $|k\ra$ topological sectors. While such contributions are irrelevant  in Minkowski space-time ${\mathbb{R}_{1,3}}$, they become important when the system is defined on certain small  compact manifolds. Without loss of generality, consider a manifold  ${\mathbb{M}}$
 which has at least one non-trivial direct factor of the fundamental group, e.g., $\pi_1[U(1)]\cong \mathbb{Z}$. The  topological sectors $|k\ra$, which play a key role in our discussions, arise precisely from the presence of such nontrivial mappings for the $U(1)$ Maxwell gauge theory. The corresponding physically observable phenomenon has been termed the topological Casimir effect (TCE). 
 
 In particular,  it has been explicitly shown in  \cite{Cao:2013na}  that these novel terms in  the topological portion of the partition function ${\cal{Z}}_{\rm top}$  lead to a fundamentally new  contribution to the Casimir vacuum pressure that appears as a result  of tunneling events between topological sectors $|k\ra$. 
 Furthermore,   ${\cal{Z}}_{\rm top}$  displays  many features 
 of   topologically ordered systems, which were  initially   introduced in the context of condensed matter (CM) systems (see recent reviews \cite{Cho:2010rk,Wen:2012hm,Sachdev:2012dq, Cortijo:2011aa, Volovik:2011kg}):  ${\cal{Z}}_{\rm top}$ demonstrates  the degeneracy of the system which can only be described in terms of non-local operators  \cite{Zhitnitsky:2013hba}; the infrared physics of the system can be studied  in terms of  non-propagating auxiliary topological fields \cite{Zhitnitsky:2014dra}, analogous to how a topologically ordered system  can be  analyzed  in terms of  the Berry's connection (also an emergent rather than fundamental field), and the corresponding expectation value of the  auxiliary topological field  determines the phase of the system.  In fact,  this technical trick of describing the system in terms of auxiliary fields will play a key role in our present discussions.  
   
 As we review in section \ref{magnetic},  the relevant   vacuum fluctuations  which saturate the topological portion of the partition function ${\cal{Z}}_{\rm top}$ are formulated in terms of topologically nontrivial boundary conditions. Classical instantons formulated in Euclidean space-time satisfy the periodic boundary conditions up to a large gauge transformation and  provide topological magnetic instanton fluxes in the $z$-direction.
 These integer magnetic fluxes describe the tunneling transitions between physically identical but topologically distinct $|k\ra$ sectors. 
 Precisely these field configurations generate   an  extra Casimir vacuum pressure in the system.
 
 What happens to this complicated vacuum structure when the system is placed in the background of a constant external magnetic field  $B_{\rm ext}^z$? The answer is known \cite{Cao:2013na}: the corresponding partition function ${\cal{Z}}_{\rm top}$ as well as all  observables, including the topological contribution to the Casimir pressure,  are highly sensitive to small magnetic fields and demonstrate $2\pi$ periodicity  with respect to the external magnetic flux represented by the parameter $\theta_{\rm eff}\equiv eSB_{\rm ext}^z$  where $S$ is the $xy$ area   of the system  ${\mathbb{M}}$.  This sensitivity to external magnetic field  is a result of  the quantum interference  of the external field with the topological quantum fluctuations. Alternatively, one can see this as resulting from a small but non-trivial overlap between the conventional Fock states, constructed by perturbative expansions around each $|k\rangle$ sector, and the true energy eigenstates of the theory, which are only attainable in a non-perturbative computation that takes the tunneling into account. This strong ``quantum" sensitivity of the TCE should be contrasted  with conventional Casimir forces  which are  practically unaltered by any external    field due to the strong suppression $\sim B_{\rm ext}^2/m_e^4$ (see   \cite{Cao:2013na} for the details).

What happens when the external E\&M field depends on time? It has been argued in \cite{Zhitnitsky:2015fpa,Cao:2015uza}  that the corresponding systems will radiate real physical photons with transverse polarizations. However, the arguments of Ref.  \cite{Zhitnitsky:2015fpa,Cao:2015uza} were based on purely classical considerations at   small frequencies $\omega\rightarrow 0$ of the external fields. 
The main goal of the present work is  to the study  the quantum dynamics of the topological    vacuum  transitions between $|k\rangle$ states
 in the presence of a {\it rapidly time-varying}   external  E\&M field. 
 
The fundamental technical difficulty  for such an analysis is that the radiation of real physical particles on mass shell is inherently formulated in {\it Minkowski}  space-time with a well-defined Hilbert space of asymptotic states. At the same time, the vacuum fluctuations (``instanton fluxes")  interpolating between the topological   $|k\rangle$  sectors  and  saturating  the path integral are  fundamentally formulated in {\it Euclidean} space-time\footnote{\label{QCD}This problem is not  specific to our system. Rather, it is a quite common problem when the path integrals are performed in Euclidean space-time, but the relevant physical questions are  formulated in   Minkowski terms. In particular, the problem is well known in QCD lattice simulations  with  conventional Euclidean formulations. All questions on non-equilibrium  dynamics and particle production represent challenges for the QCD lattice community.}.
  
We overcome this obstacle by introducing auxiliary topological fields to effectively describe the tunneling transitions computed in Euclidean space-time. These auxiliary fields can be analytically continued to Minkowski space-time.  After making the connection between the  auxiliary topological fields and the Minkowski observables, we proceed using conventional Minkowski-based  techniques, including the construction of the creation and annihilation operators, coherent states,  the appropriate Hamiltonian   describing the coupling of the microwave cavity to the system,  etc. 
  
  Our presentation is organized as follows. In Section \ref{review}, we review the relevant elements of the system including the formulation of the magnetic (\ref{magnetic}) and electric (\ref{electric})  instanton fluxes. In Section \ref{dipole moment operators}, we construct the dipole moment operators (electric and magnetic types) using our auxiliary fields continued to Minkowski space-time. In section \ref{metastable}, we formulate the problem of radiation  in proper quantum mechanical terms by identifying the quantum ``states" of the system and studying the quantum matrix elements  between them. 
   In Section \ref{q-transitions},  we discuss quantum transitions in the system in a cavity in the presence of a time-dependent external E\&M field. In the concluding Section \ref{conclusion}, we speculate that 
  the same   ``non-dispersive"  type of vacuum energy (which cannot be  expressed in terms of any propagating degrees of freedom, and which is the subject of the present work) might be responsible for the de Sitter phase of our Universe, where the vacuum energy plays a crucial  role in its evolution.

  \section{Topological partition function. Euclidean path integral formulation}\label{review}
  Our goal here is to review the Maxwell system  defined  on a Euclidean  4-manifold $\mathbb{I}^1\times \mathbb{I}^1\times \mathbb{S}^1\times \mathbb{S}^1$       with  sizes $L_1 \times L_2 \times L_3 \times \beta$ in the respective directions. This construction  provides the infrared regularization of the system, which plays a key role in the proper treatment of the  topological terms  related to tunneling events between topologically distinct but physically identical $|k\ra$ sectors. We start in section \ref{magnetic} with the construction of the magnetic  instanton fluxes considered  in \cite{Cao:2013na} and continue  in Section \ref{electric} with  the electric instanton fluxes considered in \cite{Cao:2015uza}. The construction of the respective instantons (\ref{toppot4d}) and (\ref{A_top}) have been discussed in the earlier works \cite{Cao:2013na,Cao:2015uza} and even earlier in the original studies of the Schwinger model in 2d \cite{SW,Azakov}, so we leave a review of the relevant  details to Appendix \ref{appendix_instantons}. Discussions on how these instanton fluxes can be generated in experiment with suitable boundary conditions can be found in Appendix \ref{design}.
  
  \subsection{Magnetic type instantons}\label{magnetic}
  In what follows we simplify our analysis by considering   a clear case with topological winding sectors $|k\ra$    in the $z$-direction only.    This simplification can be justified with the  geometry $L_1, L_2 \gg L_3 , \beta$, similar to the construction of the conventional CE.   
  In this case, our system resembles the 2d Maxwell theory in  \cite{Cao:2013na} by dimensional reduction: taking a slice of the 4d system in the $xy$-plane will yield precisely the topological features of the 2d torus.
 With this geometry, the dominant classical instanton configurations that describe tunneling transitions can be written as
  \be
  \label{toppot4d}
  A^{\mu}_{\rm top} = \left(0 ,~ -\frac{\pi k}{e L_{1} L_{2}} x_2 ,~ \frac{\pi k}{e L_{1} L_{2}} x_1 ,~ 0 \right),
  \ee  
  where $k$ is the winding number that labels the topological sector. \exclude{The terminology ``instanton" is adapted   from  similar 
  studies in 2d QED    \cite{Cao:2013na} where the corresponding configuration in the $A_0=0$ gauge describes the interpolation between pure gauge vacuum winding states $|k\ra$. We  use the same terminology and interpretation for the 4d case  because (\ref{topB4d}) is the classical configuration saturating the partition function ${\cal{Z}}_{\rm top}$, in close analogy with the 2d case (details in \cite{Cao:2013na}).} 

This classical instanton  configuration satisfies the periodic boundary conditions up to a large gauge transformation,  and provides a topological magnetic instanton flux in the $z$-direction:
  \be
  \label{topB4d}
  \vec{B}_{\rm top} &=& \vec{\nabla} \times \vec{A}_{\rm top} = \left(0 ,~ 0,~ \frac{2 \pi k}{e L_{1} L_{2}} \right),\\
  \Phi&=&e\int dx_1dx_2  {B}_{\rm top}^z={2\pi}k. \nonumber
  \ee
  The Euclidean action of the system is quadratic and has the  form  
  \be
  \label{action4d}
  \frac{1}{2} \int \dd^4 x \left\{  \vec{E}^2 +  \left(\vec{B} + \vec{B}_{\rm top} +\vec{B}_{\rm ext}\right)^2 \right\} ,
  \ee
  where $\vec{E}$ and $\vec{B}$ are the dynamical quantum fluctuations of the gauge field, and $\vec{B}_{\rm ext}$ is classical external magnetic field.

  As discussed in detail in \cite{Cao:2013na}, the quantum fluctuations of the gauge field decouples from the topological and external fields, allowing us to arrive at a simple expression for the topological partition function
  \be 
  \label{Z_eff}
  {\cal{Z}}_{\rm top}(\tau, \theta_{\rm eff})
  =\sqrt{\pi\tau} \sum_{k \in \mathbb{Z}} \exp\left[-\pi^2\tau \left(k+\frac{\theta_{\rm eff}}{2\pi}\right)^2\right],~~
  \ee
  where   
  \be
  \label{tau}
  \tau \equiv {2 \beta L_3}/{e^2 L_1 L_2}.
  \ee
  is a dimensionless system size parameter, and the effective theta parameter $\theta_{\rm eff} \equiv e L_1L_2 B^z_{\rm ext}$ is defined in terms of the external magnetic field $B^z_{\rm ext}$.
  Applying the Poisson summation formula  leads to the dual expression
  \be 
  \label{Z_dual1}
  {\cal{Z}}_{\rm top}(\tau, \theta_{\rm eff})
  = \sum_{n\in \mathbb{Z}} \exp\left[-\frac{n^2}{\tau}+in\cdot\theta_{\rm eff}\right]. 
  \ee
  Eq. (\ref{Z_dual1})  justifies our notation for  the effective theta parameter $\theta_{\rm eff}$ as it enters the partition function in combination with integer $n$. One should emphasize that the $n$ in the dual representation (\ref{Z_dual1}) is not the integer magnetic flux $k$ defined in Eq. 
  (\ref{topB4d}). Furthermore,  the $\theta_{\rm eff}$ parameter which enters (\ref{Z_eff}, \ref{Z_dual1}) is not the fundamental $\theta$ parameter normally introduced into the Lagrangian  in front of the $\vec{E}\cdot\vec{B}$ operator. Rather,  $\theta_{\rm eff}$ should be understood as an effective parameter representing the construction of the  $|\theta_{\rm eff}\ra$ state for each slice with non-trivial $\pi_1[U(1)]$ in the 4d system. In fact, there are three such  $\theta_{\rm eff}^{M_i}$  parameters representing different slices of the 4-torus and their corresponding external magnetic fluxes. There are similarly three $\theta_{\rm eff}^{E_i}$
  parameters representing the external electric fluxes (in Euclidean space-time)   as discussed  in \cite{Zhitnitsky:2013hba}, such that the total number of $\theta$ parameters classifying the system is six, in agreement with the total number of hyperplanes in four dimensions\footnote{Since it is not possible to have a 3D spatial torus without embedding it in 4D spatial space, the corresponding construction where all six possible types of  fluxes are generated represents a pure academic interest.\label{torus}}.

  To study the magnetic response of the system under the influence of an external magnetic field, we differentiate with respect to the external magnetic field to obtain the induced magnetic field
  \begin{align}
  & \la B_{\rm ind}\ra =-\frac{1}{\beta V} \frac{\partial \ln {\cal Z}_{\rm top}}{\partial B_{\rm ext}} =-\frac{e}{\beta L_3} \frac{\partial \ln {\cal Z}_{\rm top}}{\partial \theta_{\rm eff}} \\
  & =\frac{\sqrt{\tau \pi}}{{\cal Z}_{\rm top}} \sum_{k \in \mathbb{Z}} (B_{\rm ext} +\frac{2 \pi k}{e L_1 L_2}) \exp[-\tau \pi^2(k+\frac{\theta_{\rm eff}}{2\pi})^2].\nonumber
  \end{align}
  This induced magnetic field can also be interpreted as a magnetic dipole moment
  \begin{align}
  \label{mag_moment}
  & \la m_{\rm ind}\ra=-\la B_{\rm ind}\ra L_1 L_2 L_3 \\
  &=- \frac{\sqrt{\tau \pi}}{{\cal Z}_{\rm top}} L_3 \sum_{k \in \mathbb{Z}} \frac{\theta_{\rm eff}+2 \pi  k}{e} \exp[-\tau \pi^2(k+\frac{\theta_{\rm eff}}{2\pi})^2]. \nonumber
  \end{align}
  
  \subsection{Electric type instantons} \label{electric}
 To study the electric instanton fluxes, we consider   two parallel conducting plates which form the boundary in the $z$-direction, endowing the system with the geometry of a small quantum capacitor that has plate area $L_1 \times L_2$ and separation $L_3$ at an ambient temperature of $T=1/\beta$. \exclude{This parallel plate description still confuses me, and chances are, it will confuse the readers as well..I mean we have periodicity in the z-direction, the introduction of plates/capacitor is quite misleading in the way that it is not strictly a capacitor...it is two plates with a wire connecting them to enforce the PBC. If we use the simpler setup proposed above, then I think it is more clear to remove the capacitor description. Similar arguments apply to the title... In this section we do not discuss the question  on possible practical realizations of relevant geometry when  nontrivial mapping  $\pi_1[U(1)]$ with electric fluxes may occur.  Similar question on    nontrivial magnetic  mapping $\pi_1[U(1)]$ was discussed   in ref. \cite{Zhitnitsky:2015fpa} where it was  argued that  a simple  geometry in form of a ring is sufficient to  generate   nontrivial topological magnetic fluxes along $z$ direction with a single classification parameter $\theta_{\rm eff}$. We postpone the  corresponding discussions for the electric case until section \ref{numerics}. } These two plates are connected by an external  wire to enforce the periodic boundary conditions (up to large gauge transformations) in the $z$-direction, and so the system can be viewed as a quantum LC circuit where the external wire forms an inductor L. The quantum vacuum  between the plates (where the tunneling transitions occur) represents  the  object of our studies.

  The classical instanton configuration in Euclidean space-time which describes tunneling transitions between the topological sectors $|k\ra$ can be represented as follows:
  \be
  \label{A_top}
  A^{\mu}_{\rm top} (t) &=& \left(0 ,~ 0,~ 0, ~ \frac{2\pi k}{e L_{3} \beta } t  \right) \\  \nonumber
  A^3_{\rm top}(\beta)&=&A^3_{\rm top}(0)+\frac{2\pi k}{eL_3},
  \ee  
  where $k$ is the winding number that labels the topological sector and $t$ is the Euclidean time.  This classical instanton  configuration satisfies the periodic boundary conditions up to a large gauge transformation,  and produces a topological electric instanton flux in the $z$-direction:
  \be
  \label{E_top}
  \vec{E}_{\rm top} =\dot{\vec{A}}_{\rm top} = \left(0 ,~ 0,~ \frac{2 \pi k}{e L_{3} \beta} \right).
  \ee
  This  construction of these electric-type instantons is in fact much closer (in  comparison with the magnetic instantons  reviewed  in the previous Section \ref{magnetic}) to the  Schwinger model on a circle
  where the relevant instanton configurations were originally constructed \cite{SW,Azakov}. 
  The Euclidean action of the system takes the form 
  \be
  \label{action_E}
  \frac{1}{2} \int \dd^4 x \left\{  \left(\vec{E} + \vec{E}_{\rm top} +\vec{E}_{\rm ext} \right)^2  +  \vec{B}^2 \right\} ,
  \ee
  where, as in the magnetic case, $\vec{E}$ and $\vec{B}$ are the dynamical quantum fluctuations of the gauge field, $\vec{E}_{\rm top}$ is the topological instanton field and $\vec{E}_{\rm ext}$ is a classical external field. 
  
  Unlike magnetic fields, which remain the same under analytic continuation between Euclidean and Minkowski space-times, an electric field acquires an additional factor of $i$ as it involves the zeroth component of four-vectors,  i.e. $E_z=\partial_0 A_z - \partial_z A_0$. A detailed treatment is given in \cite{Cao:2015uza}, and here we only state the final expressions for the partition function:
\be 
\label{Z_E_theta}
{\cal{Z}}_{\rm top}(\eta, \theta_{\rm eff}^E)
=  \sum_{k \in \mathbb{Z}} \exp\left[-\pi^2\eta \left(k+\frac{\theta_{\rm eff}^E}{2\pi}\right)^2~\right],~~
\ee
for an Euclidean source $ \theta_{\rm eff}^E$, and
\be 
\label{Z_M}
\bar{{\cal{Z}}}_{\rm top}(\eta, \theta_{\rm eff}^M)
=   \sum_{k \in \mathbb{Z}} \exp{\left[-\eta\left(\pi^2 k^2+i  \pi k \theta_{\rm eff}^M   \right)\right]},~~
\ee
for a Minkowski source
\be
\label{theta_M}
\theta_{\rm eff}^M=eL_3\beta E^{\rm Mink}_{\rm ext}=-i\theta_{\rm eff}^E.
\ee
We have used the dimensionless system size parameter 
  \be
  \label{eta}
  \eta\equiv \frac{2L_1L_2}{e^2\beta L_3}. 
  \ee

Our interpretation in this case remains the same: in the presence of a physical external electric field $E^{\rm Mink}_{\rm ext}$ represented by the complex source ${\theta_{\rm eff}^E}$, the path integral (\ref{Z_E_theta}) is saturated by the Euclidean configurations (\ref{E_top}) describing physical tunneling events between the topological sectors $|k\ra$.

Now, one can compute the induced Minkowski-space electric field and dipole moment  in response to the external source $\theta_{\rm eff}^M$ by differentiating the partition function  (\ref{Z_M}) with respect to $E_{\rm ext}^{\rm Mink}$:
\begin{align}
\label{E_ind_M}
& \langle E_{\rm ind}^{\rm Mink} \rangle = -\frac 1 {\beta V}\frac{\partial \ln \bar{\mathcal{Z}}_{\rm top}}{\partial E_{\rm ext}^{\rm Mink}}=
-\frac{e}{L_1 L_2}\frac{\partial \ln\bar{\mathcal{Z}}_{\rm top}}{\partial\theta_{\rm eff}^M}\\
&= \frac{1}{\bar{\mathcal{Z}}_{\rm top}}\sum_{k\in\mathbb{Z}}\frac{2 \pi k}{ eL_3\beta} e^{-\eta\pi^2k^2}\sin \left[\pi k \eta \theta_{\rm eff}^M \right]. \nonumber 
\end{align}
The expectation value for the electric dipole moment can be competed in complete analogy with magnetic case (\ref{mag_moment}), and it is given by  
\begin{align}
\label{p_ind_M}
& \langle p_{\rm ind}^{\rm Mink}  \rangle =-\langle E_{\rm ind}^{\rm Mink} \rangle L_1 L_2 L_3 \\
&= -\frac{1}{\bar{\cal Z}_{\rm top}}\sum_{k \in \mathbb{Z}} \frac{2\pi k L_1 L_2}{e \beta} e^{-\eta \pi^2 k^2} \sin (\pi k \eta \theta^M_{\rm eff}).  \nonumber
\end{align}

\subsection{Classical dipole radiation}\label{radiation}

Although (\ref{mag_moment}) and (\ref{p_ind_M}) have been derived assuming static external magnetic and electric fields, these expressions still hold when the external fields vary slowly compared to all relevant time scales of the system. In this case, the corresponding dipole moments $\langle {m}_{\rm ind}(t)\rangle$ and $\langle {p}^{\rm Mink}_{\rm ind}(t)\rangle$ also take on time dependence in response to semiclassical time-dependent external sources as (\ref{mag_moment}) and (\ref{p_ind_M}) suggest. Hence, one can invoke the laws of classical electrodynamics to study the magnetic and electric dipole radiation as a result of this time dependence. The radiation intensity is given by the classical expressions 
\be
\label{S}
dI^{\rm M}(t)= \langle \ddot{m}_{\rm ind} (t)\rangle^2 \frac{\sin^2 \theta}{16\pi^2 c^3}d\Omega,\nonumber\\
dI^{\rm E}(t)=  \langle \ddot{p}^{\rm Mink}_{\rm ind} (t)\rangle^2 \frac{\sin^2 \theta}{16\pi^2 c^3}d\Omega,
\ee
while  the  total radiated power assumes the classical form
\be
\label{intensity}
I^{\rm M}(t)=\frac{1}{6\pi c^3 } \langle \ddot{m}_{\rm ind} (t)\rangle^2, 
~~ I^{\rm E} (t)=\frac{1}{6\pi c^3 } \langle \ddot{p}^{\rm Mink}_{\rm ind} (t)\rangle^2 ~~
\ee
for the magnetic and electric systems respectively. If one is to compute the average   intensity $\langle I(t)\rangle$ over  a cycle  assuming conventional periodic oscillation
$\sim \cos(\omega t)$ for the field, one gets
\be
\label{intensity1}
\langle I^{\rm M}\rangle=\frac{\omega^4}{12\pi c^3 } \langle  {m}_{\rm ind} \rangle^2, 
~~ \langle I^{\rm E}\rangle=\frac{\omega^4}{12\pi c^3 } \langle  {p}^{\rm Mink}_{\rm ind}\rangle^2. ~~
\ee

\begin{figure}
	\centering
	\includegraphics[width=0.55\textwidth]{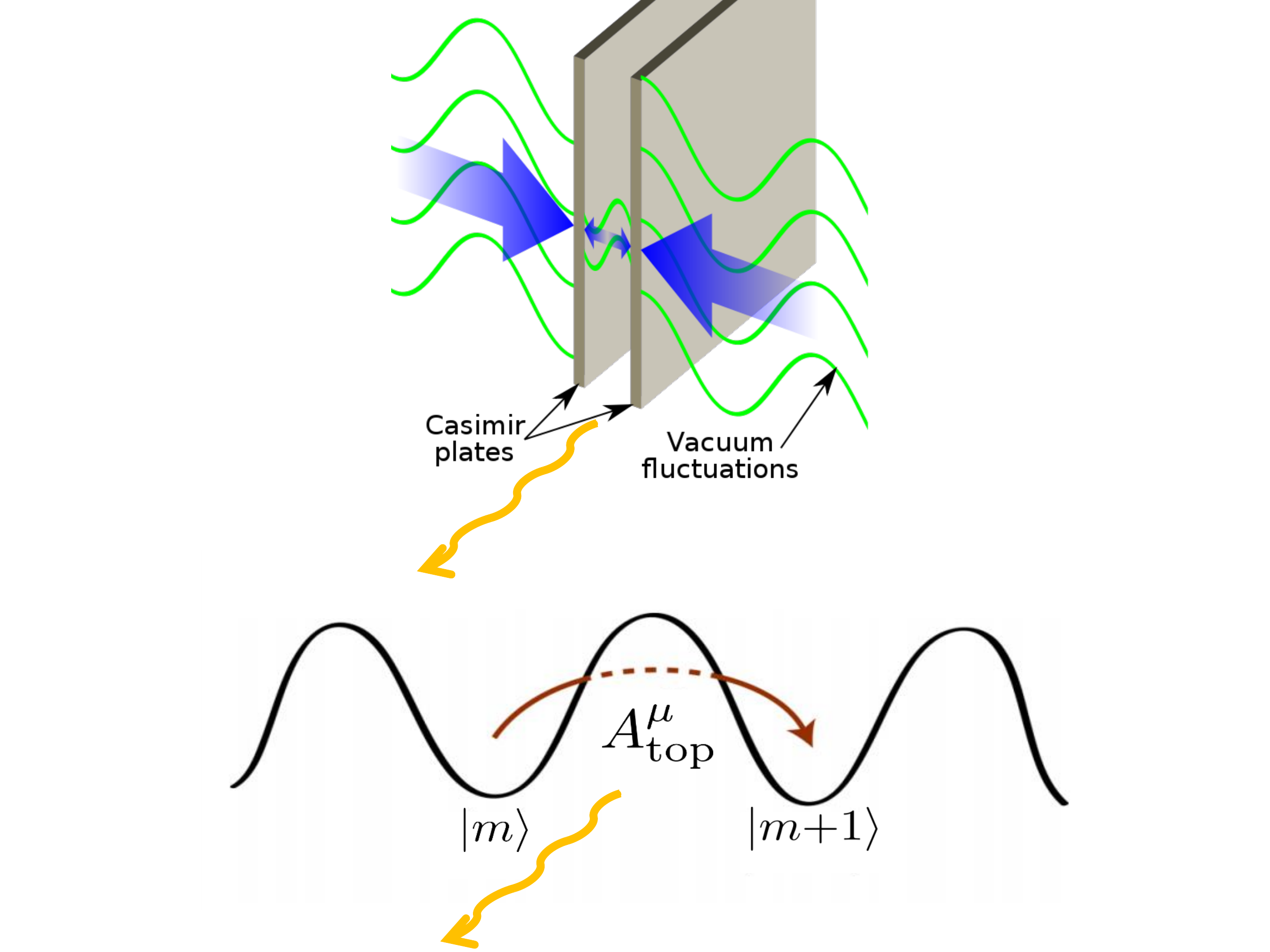}
	\caption{\label{fig:emission} An illustration of the mechanism of photon emission in the conventional DCE (top), and in the TCE, or Maxwell system on a compact manifold (bottom). In the conventional DCE, the accelerating Casimir plates turn some of the virtual photons into real on-shell propagating photons which leave the system. In the case of the TCE, the tunneling transitions between infinitely degenerate vacuum winding states $|m\ra$ are represented by instanton solutions $A_{\rm top}^\mu$.  These instanton configurations cannot be expressed in terms of physical transverse propagating E\&M fields. Precisely these topological configurations are eventually responsible for the emission of real photons in a time-dependent background, see Sections \ref{metastable} and \ref{q-transitions} for details.  }
\end{figure}

A few comments are in order. Firstly, (\ref{S}) and (\ref{intensity}) makes the important statement that the system   emits physical photons from the vacuum in the presence of time-dependent external fields, in close analogy with the dynamical Casimir effect (DCE).  Its difference from the conventional DCE \cite{DCE,DCE-review,DCE-exp} is that the radiation from  the vacuum in our system is not due to the conversion of virtual to real photons, as illustrated in the top panel of Fig. \ref{fig:emission}. Rather, it  occurs as a result  of  tunneling  events between topologically different but physically identical  vacuum winding states in a time-dependent background, and the physical photons here are emitted from these instanton-like configurations describing the tunneling transitions as illustrated in the bottom panel of Fig. \ref{fig:emission}.

 Secondly, the  magnetic  dipole radiation $\sim \langle \ddot{m}_{\rm ind} (t)\rangle^2$ can be easily understood in terms of topological non-dissipating  currents  flowing along the ring \cite{Zhitnitsky:2015fpa}, while the electric dipole radiation  $\langle \ddot{p}^{\rm Mink}_{\rm ind} (t)\rangle^2 $ can be understood in terms of fluctuating surface charges on the capacitor plates \cite{Cao:2015uza}. When the external field  fluctuates, the induced non-dissipating currents and surface charges follow suit.  This obviously leads to the radiation of real photons  as formulae (\ref{S}),  (\ref{intensity})  imply,  which we call the non-stationary TCE.  
One should emphasize that     the interpretation  of the TCE (as well as non-stationary TCE, which is the subject of the present work) in terms of topological non-dissipating currents and topological surface charges is the consequential, rather than fundamental, explanation.  The fundamental explanation is still the instantons tunneling between the topological sectors, which occur in the system even when topological boundary currents and charges  are not generated (for example, in the absence of external fields).

Finally, one should note that the above analysis of dipole radiation is purely classical: the induced dipole moments (\ref{mag_moment}) and (\ref{p_ind_M}) are  treated as classical dipoles and then varied in the semiclassical limit (such that the expressions  (\ref{mag_moment}) and (\ref{p_ind_M}) remain valid) to yield electromagnetic radiation. 

The new contribution of this paper will be presented in the following sections, where we develop quantum mechanical machinery with which to study  the emission of photons from the topological vacuum ($\cal{TV}$). This goal calls for a transition from the classical description (\ref{S}),  (\ref{intensity})   of emission  in terms of  dipole expectation values 
$ \langle {m}_{\rm ind} (t)\rangle$ and  $\langle  {p}^{\rm Mink}_{\rm ind} (t)\rangle$  to a Minkowski description  based on quantum mechanical operators, quantum states,  and transition matrix elements.
We already mentioned the  fundamental obstacle  in developing such a technique, see Footnote \ref{QCD} and the corresponding paragraph. Formula (\ref{S}),  (\ref{intensity}) will serve as the consistency check between the  classical and quantum  descriptions: it will provide some confidence that  the  quantum mechanical description (based on auxiliary topological fields developed in the next sections)  reproduces the classical formulae (\ref{S}),  (\ref{intensity})   in the low frequency limit as it should according to the correspondence principle.

\section{Dipole moment operators}
\label{dipole moment operators}
In this section, we use auxiliary fields to construct dipole moment operators  in terms of quantum mechanical operators.
 These   operators   can be analytically continued to Minkowski space-time.   They  will play a  crucial role in Section \ref{q-transitions} where we study the quantum  transitions in the system using  quantum mechanical Hilbert states formulated in Minkowski   space-time. 
 
The expectation values for the induced electric field and dipole moment in Eq. (\ref{E_ind_M}) and (\ref{p_ind_M}) were calculated in the Euclidean path integral approach at nonzero temperature $\beta^{-1}$. In what follows we  wish to formulate the topological features of our system using topological auxiliary fields and topological action.
This technique is well known to the particle physics and CM communities. In particular, it was exploited   in  \cite{BF} for the Higgs model in CM context and in \cite{Zhitnitsky:2013hs} for the so-called weakly coupled ``deformed QCD". In the present context of the Maxwell system, this technique was developed in \cite{Zhitnitsky:2014dra}, and we follow the notations from that paper.

We first illustrate how to obtain the dipole moment operator for the magnetic system reviewed in Section \ref{magnetic}, as it avoids the potentially confusing analytic continuation between Minkowski and Euclidean space-times. The same procedure can then be easily applied to the electric case reviewed in Section \ref{electric}.

\subsection{Magnetic dipole moment}\label{magnetic-dipole}
We follow  \cite{Zhitnitsky:2014dra} and 
  insert  in the original path integral (\ref{Z_eff}) the following delta functional:
\begin{align}
\label{delta_B}
&\delta [B^z - \epsilon^{zjk} \partial_j a_k(x)] \\
&\sim 
\int {\cal D}b_z \exp{\left[i \int \dd^4 x b_z(x) (B^z - \epsilon^{zjk} \partial_j a_k(x)) \right] }, \nonumber
\end{align}
where $j,k=1,2$. Here, $B^z$ is treated as the original magnetic field operator entering the action (\ref{action4d}), including both the classical $k$-instantons and the quantum fluctuations around them. Therefore, we treat $B^z$ as fast degrees of freedom. In comparison, the auxiliary fields $a_k(x)$ and $b_z(x)$ should be considered slow-varying external sources that effectively describe the large distance physics which results from tunneling transitions. We proceed by summing over all instanton configurations as before and integrating out the original fast degrees of freedom in the presence of the slow fields $a_k(x)$ and $b_z(x)$. The effective Lagrangian can then be expressed in terms of these auxiliary fields. 
 
Fortunately, the derivations can be performed as before since  the Lagrange multiplier  field $b_z(x)$ enters (\ref{delta_B}) in exactly the same manner that the external magnetic field $B_{\rm ext}\sim \theta_{\rm eff}$ enters the action  (\ref{action4d}). Therefore, we arrive at
\begin{align}
\label{Z_auxiliary}
& {\cal{Z}}_{\rm top}= \sum_{k \in \mathbb{Z}} \sqrt{\pi \tau} \int {\cal D} a {\cal D} b_z \exp{\left[ - \int \dd^4 x {\cal L} \right]} \\
& {\cal L}=\frac{2 \pi^2}{e^2 L_1^2 L_2^2} (k + \frac{\theta_{\rm eff} -i \phi(x)}{2\pi} )^2 + {\cal L}_{\rm top} \nonumber \\
&  {\cal L}_{\rm top} =i b_z(x) \epsilon^{zjk} \partial_j a_k(x) , \nonumber
\end{align}
where $\phi(x) \equiv e L_1 L_2 b_z(x)$. One can   see from (\ref{Z_auxiliary})   that the topological term $ {\cal L}_{\rm top}$  is explicitly generated 
in this effective description. This term has Chern-Simons structure which normally appears in many similar CM commutations (see e.g. \cite{ChenLee,BF}), and  one should therefore anticipate a number of topological phenomena as a result of this Chern-Simons structure. 
Furthermore, one can show \cite{Zhitnitsky:2014dra} that the auxiliary   field $ a_i(x)$ written in momentum space  $ {a_i}(\mathbf{k})$ strongly   resembles Berry's connection ${\cal{A}}_i(\mathbf{k}) $ in CM physics\footnote{ \label{Berry}In fact, one can argue that the auxiliary fields in our framework play the same role as  Berry's connection in CM physics. In particular,  as it is known Berry's phase in CM systems  effectively describes the variation of the $\theta$ parameter $\theta\rightarrow \theta-2\pi P$ as a result of the coherent influence of strongly interacting fermions that polarize the system, i.e., $P=\pm 1/2$, see e.g. \cite{ChenLee}. 
Our auxiliary fields essentially describe  the same physics. Therefore, it is not a surprise that the induced dipole moment $m_{\rm ind }$, to be discussed below, can be explicitly expressed in terms of these auxiliary fields. }.

The  integration  over $b_z$ is Gaussian, and  can be explicitly executed with the result:
\begin{align}
\label{Z_auxiliary1}
&{\cal{Z}}_{\rm top}= \sum_{k \in \mathbb{Z}} \sqrt{\pi \tau} \int {\cal D} a \exp{\left[ - \int \dd^4 x {\cal L} \right]} \\
&{\cal L}=-\frac{1}{2} \left(\epsilon^{zjk} \partial_j a_k(x)\right)^2 + \frac{2 \pi k + \theta_{\rm eff}}{e L_1 L_2} \left( \epsilon^{zjk} \partial_j a_k(x)\right). \nonumber
\end{align}
A few comments are in order.  Firstly, the negative sign in Eq. (\ref{Z_auxiliary1}) should not be considered as any inconsistency or violation of unitarity. Indeed, the  field $a_k(x)$ is an auxiliary non-propagating field  introduced into the system to simplify the analysis, and any observable could be computed without it. Instead, this field should be considered  as a saddle    point  saturating the Euclidean partition function ${\cal{Z}}_{\rm top}$ in  the path integral approach\footnote{\label{ghost}In many respects this negative sign in Eq. (\ref{Z_auxiliary1}) resembles the negative sign for the so-called Veneziano ghost in the course of the resolution of the $U(1)_A$ problem in QCD, see \cite{Zhitnitsky:2013hs}  for references and details in the given context. One can explicitly see from the computations  in \cite{Zhitnitsky:2013hs} how the negative kinetic term for the Veneziano ghost  is generated due to tunneling transitions between different topological sectors in very much the same way as it occurs in our system  represented by the effective Lagrangian (\ref{Z_auxiliary1}). Precisely this ``wrong sign" in the effective Lagrangian might be a key element in understanding the new type of cosmological vacuum energy known as the dark energy, see comments in the concluding section \ref{conclusion}.}.

Secondly, the $[-\epsilon^{zjk} \partial_j a_k(x)]$ term in the above Lagrangian couples to both the instanton field expressed in terms of  $k$ fluxes,  and the external field formulated in terms of $ \theta_{\rm eff}$. The  physical meaning of this operator can be
easily understood by noticing that it enters the  Lagrangian precisely as  how a  magnetic dipole moment density couples  to the external magnetic field.  Therefore,   we  identify $[-\epsilon^{zjk} \partial_j a_k(x) ]$ with the magnetization of the system.

To confirm this  conjecture, we should  compute the expectation value of $[-\epsilon^{zjk} \partial_j a_k(x)]$ to  reproduce the magnetic dipole moment derived in the Euclidean path integral  approach (\ref{mag_moment}). This task can be easily performed because the  integration over $ \partial_j a_k(x)$  is Gaussian and can be carried out by a conventional change of variables 
\be
\epsilon^{zjk} \partial_j a'_k(x)=\epsilon^{zjk} \partial_j a_k(x)-\frac{2 \pi k + \theta_{\rm eff}}{e L_1 L_2},
\ee 
after which the Lagrangian becomes
\be
\label{lagrangian_new}
{\cal L}=-\frac{1}{2} \left(\epsilon^{zjk} \partial_j a'_k(x)\right)^2 + \frac{1}{2} \left(\frac{2 \pi k + \theta_{\rm eff}}{e L_1 L_2}\right)^2.
\ee
The expectation value  of $[-\epsilon^{zjk} \partial_j a_k(x)]$   is then given by
\be
\label{Z_auxiliary2}
&\,& {\la m_{\rm ind}\ra}= \langle [-\epsilon^{zjk} \partial_j a_k(x)] \rangle V \\
&=& \frac{ \sum_{k \in \mathbb{Z}} \int {\cal D} a e^{-\int \dd^4 x {\cal L}} \left(-\epsilon^{zjk} \partial_j a'_k(x)-\frac{2 \pi k +\theta_{\rm eff}}{e L_1 L_2}\right)V}{ \sum_{k \in \mathbb{Z}} \int {\cal D} a e^{-\int \dd^4 x {\cal L}}} \nonumber \\
&=& -L_3\frac{\sum_{k \in \mathbb{Z}} \left(\frac{\theta_{\rm eff}+2 \pi  k}{e}\right)\exp[-\tau \pi^2(k+\frac{\theta_{\rm eff}}{2\pi})^2]}{\sum_{k \in \mathbb{Z}} \exp[-\tau \pi^2(k+\frac{\theta_{\rm eff}}{2\pi})^2]}. \nonumber
\ee
Eq. (\ref{Z_auxiliary2})  exactly reproduces our previous expectation value of the magnetic dipole moment (\ref{mag_moment}), thereby confirming the   identification  of the operator  $[-\epsilon^{zjk} \partial_j a_k(x)]$ with the magnetization of the system. 
  
We would like to mention  here  that this identification should not surprise the reader. Indeed, it has been previously argued \cite{Zhitnitsky:2014dra} that the auxiliary field can be thought of as Berry's connection\footnote{These similarities, in particular, include the following features: while  $ {a_i}(\mathbf{k})$ and  ${\cal{A}}_i(\mathbf{k})$ are gauge-dependent objects, the  observables, such as polarization or magnetization (\ref{Z_auxiliary2}) are gauge invariant (modulo $2\pi$) characteristics.    Furthermore, the main features of the systems in both cases  are formulated in terms of global  rather than local characteristics. }. 
The polarization properties of a CM  system can be computed in terms of Berry's connection ${\cal{A}}_i(\mathbf{k})$ and Berry's curvature, see Footnote \ref{Berry} with relevant references. In our case, the magnetization of the system is also expressed in terms of auxiliary fields. Therefore,  Eq. (\ref{Z_auxiliary2}) is in fact fully anticipated.

\subsection{Electric dipole moment}\label{electric-dipole}
The similar procedure can be applied to the electric system to obtain an electric dipole moment operator.
The delta functional we insert into (\ref{Z_E_theta}) is 
\begin{align}
\label{delta_E}
&\delta [E^z - \epsilon^{12jk} \partial_j a_k(x)] \\
&\sim \int {\cal D}b_z \exp{\left[i \int \dd^4 x b_z(x) (E_z - \epsilon^{12jk} \partial_j a_k(x)) \right] } ,\nonumber
\end{align}
where $j,k=0,3$, and $E^z$ is taken to be the Euclidean quantum field including the instanton configurations (\ref{E_top}) and quantum fluctuations around them.  

We follow the same procedure as before by 
integrating  out  the auxiliary field $b_z(x)$.  It  leads to the following Euclidean  Lagrangian density analogous to  Eq.   (\ref{Z_auxiliary1})  describing the magnetic case
\begin{align}
\label{Z_auxiliary3}
{\cal L} =-\frac{1}{2} (\epsilon^{12jk} \partial_j a_k(x))^2 + \frac{2 \pi k + \theta^{E}_{\rm eff}}{e \beta L_3} (\epsilon^{12jk} \partial_j a_k(x)).
\end{align}
All the comments after Eq. (\ref{Z_auxiliary1}) also  apply here for  the electric case (\ref{Z_auxiliary3}).
Furthermore, there is an additional complication for  the electric case due to the necessity for a transition to physical Minskowski space-time, i.e., we have to replace the Euclidean $\theta^E_{\rm eff}$ in (\ref{Z_auxiliary3})  by the  Minkowski expression $\theta^E_{\rm eff}$ according to relation (\ref{theta_M}):
\begin{align}
\label{Z_auxiliary4}
{\cal L} =-\frac{1}{2} (\epsilon^{12jk} \partial_j a_k(x))^2 + \frac{2 \pi k + i \theta^{M}_{\rm eff}}{e \beta L_3} (\epsilon^{12jk} \partial_j a_k(x))  ,
\end{align}
where  $\theta^{M}_{\rm eff}=e\beta L_3 E^{\rm Mink}_{\rm ext}$ represents the physical electric field.
The only difference from the magnetic case is the emergence of the factor $i$ in front of the effective theta parameter. Thus, we identify  the electric dipole moment operator in Minkowski space-time with $[-i\epsilon^{12jk} \partial_j a_k(x)]V$. In what follows we confirm this conjecture by explicit computation of the 
corresponding expectation value. 

To proceed with this task we make a shift 
\be
\epsilon^{12jk} \partial_j a'_k(x)=\epsilon^{12jk} \partial_j a_k(x)-\frac{2 \pi k +i \theta^{M}_{\rm eff}}{e L_3 \beta},
\ee 
such that  the Lagrangian (\ref{Z_auxiliary4}) in terms of the new variable  $a'_k(x)$ becomes
\be
\label{Z_auxiliary5}
{\cal L}=-\frac{1}{2} \left(\epsilon^{12jk} \partial_j a'_k(x)\right)^2 + \frac{1}{2} \left(\frac{2 \pi k +i \theta^{M}_{\rm eff}}{e L_3 \beta}\right)^2.
\ee
We can now calculate  the expectation value of the electric  dipole moment:  
\be
\label{Z_auxiliary6}
&\,& \langle p_{\rm ind}^{\rm Mink}  \rangle = \langle [-i\epsilon^{12jk} \partial_j a_k(x)] \rangle V \\
&=& \frac{ \sum_{k \in \mathbb{Z}} \int {\cal D} a e^{-\int \dd^4 x {\cal L}} \left(-i\epsilon^{12jk} \partial_j a'_k(x)-i\frac{2 \pi k}{e L_3 \beta}\right)V}{ \sum_{k \in \mathbb{Z}} \int {\cal D} a e^{-\int \dd^4 x {\cal L}}} \nonumber \\
&=&  \frac{\sum_{k \in \mathbb{Z}} \left(-i\frac{2 \pi L_1L_2 k}{e\beta} \right)\exp{\left[-\eta\left(\pi^2 k^2+i  \pi k \theta_{\rm eff}^M   \right)\right]}}{ \sum_{k \in \mathbb{Z}} \exp{\left[-\eta\left(\pi^2 k^2+i  \pi k \theta_{\rm eff}^M   \right)\right]}} \nonumber\\
&=& -\frac{1}{\bar{\cal Z}_{\rm top}}\sum_{k \in \mathbb{Z}} \frac{2\pi k L_1 L_2}{e \beta} e^{-\eta \pi^2 k^2} \sin (\pi k \eta \theta^M_{\rm eff}),\nonumber
\ee
where $\bar{\cal Z}_{\rm top}$ is defined in  (\ref{Z_M}).  Here,  we have removed the constant external term to keep only the truly induced contribution to the dipole moment, consistent with our previous definition   in Section \ref{electric}. Eq. (\ref {Z_auxiliary6})
  exactly reproduces    our previous expression (\ref{p_ind_M}) which was originally derived without even mentioning any auxiliary fields. This supports once again our formal manipulations with the auxiliary fields, and it also confirms our interpretation of the operator $[-i\epsilon^{12jk} \partial_j a_k(x)]$ as the quantum polarization operator of the system. All the comments we have made  in Section \ref{magnetic-dipole} regarding the physical meaning of this operator also  apply here to the electric case, including the connection with Berry's phase, which we will not repeat here.

To study the quantum mechanical dipole transitions, we must work in Minkowski space-time where the metric signature allows for propagating on-shell photons. Although the original derivation in this section is performed in Euclidean space-time, we claim that the dipole moment operator 
$P_z\equiv[-i\epsilon^{12jk} \partial_j a_k(x)]V $ represents an operator in Minkowski space-time, as confirmed by the explicit expectation value calculation (\ref{Z_auxiliary6}). 

Our next task is to infer from our previous   Euclidean path integral computations the  structure of the quantum states, which can then be employed for conventional quantum dipole transitions in Minkowski terms, see Footnote \ref{QCD} and the related paragraph for explanation of the source of this technical subtlety.

 \section{Metastable quantum states in the Maxwell system}\label{metastable}
 The main goal of this section is to identify quantum mechanical states in Hilbert space in Minkowski space using the operators constructed in previous section
 \ref{dipole moment operators}. These quantum states have never been explicitly constructed in the previous path integral treatment of this model  \cite{Cao:2013na,Zhitnitsky:2013hba,Zhitnitsky:2014dra,Zhitnitsky:2015fpa,Cao:2015uza}. We substantiate our identification  by reproducing   the computed transition matrix elements  with  corresponding path integral computations in Euclidean space.
  
\exclude{ 
In the previous section \ref{radiation}, we discussed what we called non-stationary TCE where physical photon are radiated from the topological vacuum ($\cal{TV}$) in the presence of a time-dependent external  E\&M field.   As the induced dipole moment $\la p^{\rm Mink}_{\rm ind}\ra$ is a function of $E_{\rm ext}^{\rm Mink}$, varying the external field naturally leads to a varying dipole moment and eventually gives rise to electromagnetic radiation according to classical electrodynamics. Although the instanton configuration and the induced dipole moment are genuine quantum entities, the treatment of radiation from an semiclassical varying dipole moment in Section  \ref{radiation} was  purely  classical. In this section, we make   one step further and  formulate   the problem in proper quantum mechanical terms    by identifying the quantum ``states'' of our system and studying the quantum dipole transitions between them.
}

Before we proceed we would like to overview a well-known formal mathematical analogy between the construction of the $|\theta\ra$ vacuum states in gauge theories and Bloch's construction of the allowed/forbidden bands  in CM physics (see e.g. \cite{Shifman}). 
The large gauge transformation operator $\cal{T}$ plays the role of the crystal translation operator in CM physics. $\cal{T}$ commutes with the Hamiltonian $H$ and changes the topological sector of the system 
 \be
 \label{large_gauge_transform}
 {\cal{T}}|m\ra=|m+1\ra, ~~ [H, {\cal{T}}]=0, 
 \ee
 such that 
   the $|\theta\ra $-vacuum state  is an eigenstate of the large gauge transformation operator $\cal{T}$:
\be
|\theta\rangle =\sum_{m\in\mathbb{Z}} e^{im \theta} |m\rangle, ~~~  {\cal{T}} |\theta\ra=e^{-i\theta}  |\theta\ra .\nonumber
\ee
The $\theta$ parameter in this construction plays the role of  the ``quasi-momentum" $\theta\rightarrow qa$ of a quasiparticle propagating in the allowed energy band in a crystal lattice with unit cell length $a$. 

An important element, which is typically skipped in presenting  this   analogy but which  plays a key role in our studies is the presence of the  Brillouin zones classified by integers $k$. Complete classification can be either presented in the so-called extended zone scheme where $-\infty < qa < +\infty$, or the reduced  zone scheme where each state is classified by two numbers, the quasi-momentum $-\pi \leq qa \leq +\pi$ and the Brillouin zone number $k$. 
 
In  the classification of the vacuum states, this corresponds to describing the system by two numbers $|\theta, k\ra$, where $\theta$  is assumed to be varied in the conventional range $\theta\in [0, 2\pi)$, while the integer $k$ describes the ground state (for $k=0$) or the excited metastable vacuum states ($k\neq 0$).  In most studies devoted to the analysis of the $\theta$ vacua, the questions  related to the metastable vacuum states have not been addressed. Nevertheless, it has been known for some time  that  the   metastable vacuum states must be present in non-abelian  gauge  systems in the large $N$ limit  \cite{Witten:1980sp}.
A similar conclusion  also  follows from the  holographic description of QCD as originally discussed in \cite{wittenflux}.  Furthermore,    the metastable  vacuum states can be explicitly constructed in a weakly coupled ``deformed QCD" model  \cite{Bhoonah:2014gpa}. 

Such metastable states  will also emerge in our  Maxwell systems defined on a compact manifold.     Thus, the complete  classification of the states in our system  is $|\theta_{\rm eff}, k\ra$, where the integer $k$ plays a role similar to  the $k$-th Brillouin zone in the reduced zone classification as we discussed  above. 

\subsection{Identification of quantum states: magnetic system}\label{magnetic-states}
Through the formal manipulation in Section \ref{magnetic-dipole} we have identified the magnetic  dipole moment operator $M_z=- \epsilon^{zjk} \partial_j a_k(x) \cdot V$. We have also seen that the quantum mechanical expectation value of $\la M_z\ra$ reproduces the  expectation value $\la m_{\rm ind}^{\rm Mink}  \ra$ computed using Euclidean path integrals (\ref{mag_moment}),   i.e.
\be
\label{M-operator}
\la M_z \ra= -\left(\frac{2\pi L_3}{e}\right)\frac{\sum_{k \in \mathbb{Z}}    k  \exp[-\tau \pi^2(k+\frac{\theta_{\rm eff}}{2\pi})^2]}{\sum_{k \in \mathbb{Z}} \exp[-\tau \pi^2(k+\frac{\theta_{\rm eff}}{2\pi})^2]}.~~
\ee
Formula (\ref{M-operator}) determines a  truly induced magnetic moment  when  the trivial constant contribution (related to the external magnetic field)
is removed from the corresponding expression (\ref{mag_moment}). Formula (\ref{M-operator}) was derived  using conventional path integrals in  Euclidean space-time  
without   interpreting it  in terms of any physical states. 

Now we interpret the result    (\ref{M-operator}) in terms of quantum mechanical states in Hilbert space. 
Firstly, the   factor $ \exp[-\tau \pi^2k^2]$ originates from the partition function $ {\cal{Z}}_{\rm top}(\tau, \theta_{\rm eff})$  (\ref{Z_eff}). This exponential form in Euclidean space  suggests that the combination 
\be
\label{energy_B_k}
\epsilon(k)=\frac{\tau \pi^2 k^2}{\beta}=\frac{2L_3\pi^2k^2}{e^2L_1L_2}
\ee
can be interpreted as the energy  of state $|k\ra$ for  $\theta_{\rm eff}=0$ in Minkowski space.  In the case of a non-zero external field $\theta_{\rm eff}\neq 0$, the corresponding energy levels get shifted accordingly as in the well-known problem for a particle on a circle, 
\be
\label{energy_B_k1}
 \epsilon(k, \theta_{\rm eff})=\frac{\tau \pi^2(k+\frac{\theta_{\rm eff}}{2\pi})^2}{\beta}.
\ee
We identify the parameter $k$ with the label of the metastable vacuum state $|k\ra$, similar to the classification of the $k$-th Brillouin zone  in   CM systems mentioned above.  This interpretation is supported by the observation that for $k=0$ the energy  $\epsilon(k=0,  \theta_{\rm eff})= \frac{1}{2}(VB_{\rm ext}^2)$  is precisely the magnetic energy of the external field, while quantum tunneling generates the  excited $|  \theta_{\rm eff}, k\ra$ states with energies (\ref{energy_B_k1}). In contrast to conventional quantum states in the context of dipole transitions, the ``states'' in  our system are the $k$-instantons that describe tunneling transitions between the infinitely many degenerate vacuum winding states.
 
Once we accept this interpretation along with the identification of the magnetic  dipole moment operator $M_z=- \epsilon^{zjk} \partial_j a_k(x) \cdot V$, we then proceed to interpret the corresponding factor  in (\ref{M-operator}) as the non-vanishing transition matrix element $\la k| M_z|0\ra$ rather than a diagonal expectation value $ \la k| M_z|k\ra$. 

To simplify notations in what follows, we consider vanishing external field   $\theta_{\rm eff}=0$ and the lowest excited metastable state $k=1$, which  can be formally achieved by considering the limit $\tau\gg 1$. In this  case we can interpret (\ref{M-operator}) as the transition matrix element between the first excited state and the ground state.  
 \be
\label{transition_element-1}
\la k=0| M_z|k=1\ra\simeq -\frac{2\pi  L_3}{e }\cdot e^{-\tau \pi^2}. 
\ee
 The main argument behind   this interpretation is the observation 
that the integer parameter $k$ which enters (\ref{M-operator})  originally appeared in the Euclidean path integrals as the instanton action describing the {\it interpolation} between two topologically distinct states according to Eq. (\ref{Z_eff}). The same interpretation also follows 
from the boundary conditions (\ref{toppot4d}) such that (\ref{transition_element-1}) can be thought of (in Minkowski terminology) as the configuration describing the transition matrix element between the states which satisfy the non-trivial boundary conditions (\ref{toppot4d}) with $k=1$ and states which satisfy the trivial boundary conditions with $k=0$. 

Yet another argument supporting the Hamiltonian interpretation  in terms of the transition matrix elements (\ref{transition_element-1}) is the successful matching of our final formula for the intensity of radiation with the classical expression   for emission (\ref{intensity1}) discussed in Section \ref{radiation}. Indeed, the conventional quantum mechanical formula for the probability for the quantum transition per unit time is known to match well  with the classical formula (\ref{intensity1})  for the intensity of radiation. This spectacular example of classical correspondence  implies that the probability for the quantum emission $R_{1\rightarrow 0}$ is expressed in terms of the transition matrix element (\ref{transition_element-1}) to match the classical formula (\ref{intensity1}) 
\be
\label{correspondence}
R_{1\rightarrow 0}=\frac{\omega^3 \mu_0}{3\pi\hbar c^3} |\la k=0| M_z|k=1\ra|^2, ~  \hbar \omega R_{1\rightarrow 0}  \rightarrow\la  I^{\rm M}\ra.~~~~~
\ee
In this well known correspondence the  magnetic moment ${m}_{\rm ind}$  as usual is   identified with  the time dependent transition matrix element ${m}_{\rm ind}(t)=[\la k=0| M_z|k=1\ra e^{-i\omega t}]$. In this case  the   magnetic moment entering formula (\ref{intensity1}) for the classical emission should be identified with $[{m}_{\rm ind}(t)+{m}^*_{\rm ind}(t)]\sim 2 \cos(\omega t)$
while the magnetic moment entering the quantum mechanical expression (\ref{correspondence}) should be identified with transition matrix element (\ref{transition_element-1}).
This well-known correspondence between classical and quantum descriptions once again 
supports our interpretation of (\ref{transition_element-1}) as the transition matrix element between the excited  and ground states, though the original computations  (\ref{M-operator})  from which formula (\ref{transition_element-1}) was inferred  were performed in the Euclidean path integral approach without  any notions of the Hamiltonian formulation. 

We conclude with the following remarks.  As we mentioned previously, the expectation value  (\ref{M-operator})  vanishes when the external field is zero, though we claim 
that the transition matrix element (which eventually leads to the emission of real photons) does not vanish according to  (\ref{transition_element-1}).
There is no contradiction here as the expectation value  (\ref{M-operator})  vanishes at $\theta_{\rm eff}=0$  as a result of cancellation between $k=\pm1 $ states, while in our discussions above  we selected a single state  $|k=1\ra$ which obviously must be somehow produced by non-equilibrium dynamics and separated from the $|k=-1\ra$ state. 

Finally, 
the transitions between the quantum states $|k=1\ra\rightarrow |k=0\ra$ described here should not be confused  with  multiple  tunneling transitions between the infinitely degenerate $|n\ra$ vacuum winding states that make up the $\theta$-vacuum, classified by two parameters $|\theta_{\rm eff}, k\ra$ as discussed at the  very beginning of this section.     Unlike the vacuum winding states, these quantum states are separated in energy (\ref{energy_B_k}) and the transitions between them form the central subject of this section.

\subsection{Identification of quantum states: electric system}\label{electric-states}

Through the formal manipulation in Section \ref{electric-dipole} we have identified the dipole moment operator $P_z=- i\epsilon^{12jk} \partial_j a_k(x) \cdot V$, whose quantum mechanical expectation value reproduces the  expectation value $\la p_{\rm ind}^{\rm Mink}  \ra$ computed in the Euclidean path integral approach, i.e.,
\be
\label{P_operator}
\la P_z\ra=-\frac{1}{\bar{\cal Z}_{\rm top}}\sum_{k \in \mathbb{Z}} \frac{2\pi k L_1 L_2}{e \beta} e^{-\eta \pi^2 k^2} \sin (\pi k \eta \theta^M_{\rm eff}).
\ee
Following the magnetic system in the previous section,  we wish to  interpret this expression in terms of quantum states in Hilbert space. In the $\theta^M_{\rm eff}=0$ limit, the energy of each state can be read off the Boltzmann factors:
\be
\label{energy_k2}
\epsilon(k)=\frac{\eta \pi^2 k^2}{\beta}=\frac{2 \pi^2 k^2 L_1 L_2}{e^2 \beta^2 L_3},
\ee
 analogous to (\ref{energy_B_k}). 
As in Section \ref{magnetic-states}, we  work in the reduced zone scheme with $\theta_{\rm eff}^M \in [-\pi,\pi]$ and identify the configurations labeled by integers $k$ as the   quantum states $|k\ra$. In particular, $k=0$ is the ground state and $k\neq 0$ represents the excited metastable states. The supporting arguments made in Section \ref{magnetic-states} apply to the electric system as well.

This connection allows us to further identify the transition elements of the $P_z$ matrix from (\ref{P_operator}) where we keep only the $k=1$ state to simplify the notations:
\be
\label{transition_element}
\la k=0| P_z|k=1\ra=-i\frac{2\pi  L_1 L_2}{e \beta} e^{-\eta (\pi^2   +i\pi   \theta^M_{\rm eff})}
\ee
which is analogous to formula (\ref{transition_element-1}) for the magnetic system. 
 
 One can repeat the arguments presented in the previous subsection \ref{magnetic-states}  to infer that the correspondence formula for the 
 electric dipole transition assumes the form
   \be
\label{correspondence1}
R_{1\rightarrow 0}=\frac{\omega^3}{3\pi\hbar c^3 \epsilon_0} |\la k=0| P_z|k=1\ra|^2, ~  \hbar \omega R_{1\rightarrow 0}  \rightarrow\la  I^{\rm E}\ra.~~~~~
\ee
This   example of classical correspondence  implies that the probability for the electric dipole transition  $R_{1\rightarrow 0}$ is expressed in terms of the transition matrix element (\ref{transition_element}) to match the classical formula (\ref{intensity1}). 

Our comments after Eq. (\ref{correspondence}) for the magnetic case still hold for the electric case, and we shall not repeat them here. 
The only additional remark we would like to make to conclude this section is as follows. All our  results on the identification of the dipole moment operators and  their expectation values (\ref{M-operator}) and (\ref{P_operator}) are based on the Euclidean path integral approach. We did not and could not construct the corresponding Hilbert space and the corresponding wave functionals $\Psi_k[{A_i}]$ in Minkowski space-time which would depend on the E\&M field configurations. However, using the correspondence principle (and some other hints and indications) we were able to reconstruct the relevant matrix elements (\ref{transition_element-1}) and (\ref{transition_element}) without complete knowledge of the  wave functionals $\Psi_k[{A_i}]$. Fortunately, this is the only information we need in our following studies of quantum dipole transitions in a cavity. 

 \section{Quantum dipole transitions in a cavity}\label{q-transitions}
The  goal of this section is to construct the effective Lagrangian describing the interaction between the physical E\&M fields and the auxiliary fields introduced in Sections \ref{dipole moment operators} and  \ref{metastable}. This coupling will allow us to carry out proper quantum computations for the rate of emission of real physical photons, because the relevant transition matrix elements  (\ref{transition_element-1}) and (\ref{transition_element}) have been computed in Minkowski space-time.
 This puts us in a position to use the well developed procedure to study quantum dipole transitions, such as in the phenomenon of stimulated emission.

 Numerically the decay rate (\ref{correspondence1}) is extremely low (see Section \ref{numerics} for numerical estimates). It has been known for quite some time that different types of microwave (optical) cavities can drastically increase the sensitivity for photon detection.   Due to the smallness of the magnitude of all the topological effects of our Maxwell system, including the intensity of photon radiation, there might be hope that the stimulated emission of photons from the capacitor configuration can be detected using  microwave (optical)  resonators. 

Essentially, we adopt the conventional  technique normally used  to study a system consisting of an  atom in an  optical cavity.  The  role of the atom in our case is played by the topological Maxwell system  as described in the previous sections, while the optical cavity is replaced by a microwave cavity as  the typical frequencies for our  system are much smaller than atomic frequencies. 

However, it should be noted that  a specific design for   microwave cavities in a possible experiment is certainly beyond the scope of this paper, and we shall proceed with only a general sketch of the possible experimental setup for illustrative purposes exclusively.  Our numerical estimates given in Section  \ref{numerics} suggest that the typical sizes  where persistent currents have been observed and where coherent Aharonov-Bohm phases can be maintained could be a good starting point for a possible design. However,  we are reluctant to put forward a specific experimental setup since our main goal is to describe a new phenomenon, rather than to design a device for its observation or measurement.    We  leave the questions on possible  design for others in the community who can then use their own expertise to devise suitable experimental apparatuses.

\subsection{Coupling with quantum E\&M field}
First, we want to demonstrate that the quantum propagating E\&M field couples to the   magnetic  and electric dipole moment operators $M_z$ and $P_z$ in exactly the same way as it does to the dipole moment operators in conventional quantum mechanics.
Indeed, from (\ref{Z_auxiliary1}) one can deduce that the interaction of the quantum field $B^{\rm quant}$ with the auxiliary fields $a_k(x)$ is given by the following extra term $\Delta{\cal L}^M_{\rm int}$ in the Lagrangian 
\be
\label{int_M}
\Delta{\cal L}^M_{\rm int}= B^{\rm quant}_z \cdot \left[\epsilon^{zjk} \partial_j a_k(x)\right]  , 
\ee
where $\vec{B}^{\rm quant}=\vec{\nabla}\times \vec{A}^{\rm quant}$ is expressed in  terms of the conventional quantum propagating  field $ \vec{A}^{\rm quant}$. 
The relation (\ref{int_M}) follows from the fact that the $\theta_{\rm eff}$ parameter  entering (\ref{Z_auxiliary1})  represents the total E\&M field, including the classical and the quantum parts, i.e.
$\theta_{\rm eff}=eL_1L_2(B^{\rm class}_z +B^{\rm quant}_z)$. In our previous discussions we kept only classical, constant,   portion of the field. In our present discussions in this section we obviously need the quantum,  fluctuating,   portion of the field as well. 

The expression (\ref{int_M}) obviously has the structure of a quantum field $B^{\rm quant}$ interacting with the magnetic moment operator expressed in terms of the auxiliary fields and   derived in  (\ref{Z_auxiliary2}) using the Euclidean path integral approach. Precisely the   matrix element of this operator has been computed in   (\ref{transition_element-1}).
The operator $M_z$ and its transition matrix element   play the same role in our computations as the electron magnetic moment operator $\vec{\mu}=\frac{e\hbar}{2mc}(\vec{l}+2\vec{s})$ and the corresponding   matrix elements do in atomic physics  with the conventional coupling $-\vec{\mu}\cdot \vec{B}$. 

The same arguments also apply to the quantum coupling   of the E\&M quantum  field with the electric dipole moment operator $P_z$. Indeed, from (\ref{Z_auxiliary4})
one can deduce that the interaction of the quantum field $E^{\rm quant}$ with the auxiliary field $a_k(x)$ is given by the following extra term $\Delta{\cal L}^E_{\rm int}$ in the Lagrangian 
\be
\label{int_E}
\Delta{\cal L}^E_{\rm int}= E^{\rm quant}_z\cdot \left[ i\epsilon^{12jk} \partial_j a_k(x)\right]  .
\ee
This is because  $\theta^M_{\rm eff}$ which enters (\ref{Z_auxiliary4})
represents the physical electric field, including the constant external part and the  fluctuating quantum part.  
The expression (\ref{int_E}) obviously has the structure of the interaction between the quantum field $E^{\rm quant}$  and an electric dipole  operator expressed in terms of the auxiliary fields and   derived in  (\ref{Z_auxiliary6}) using the Euclidean path integral approach. Precisely the   matrix element of this operator has been computed in the previous section (\ref{transition_element}).
The operator $P_z$ and its transition matrix element   play the same role as the electron dipole moment operator $ \vec{d}=e\vec{r}$ and the corresponding   matrix elements do in atomic physics with conventional coupling $-\vec{d}\cdot \vec{E}$. 

The essence of the auxiliary fields $a_k(x)$ employed above  is that they   effectively account for the interaction between nontrivial topological configurations (which themselves describe the tunneling events) and the propagating physical photons.  All the relevant information about these auxiliary fields, originally introduced  in the Euclidean path integral approach,  is encoded  now in terms of the   matrix elements (\ref{transition_element-1}) and (\ref{transition_element}) 
in Minkowski space-time such that one  can proceed with the  computations of the quantum transitions using conventional Hamiltonian  techniques, which we shall do in the next section. 

\subsection{Jaynes-Cummings Hamiltonian for the topological Maxwell system}
We consider the electric system and limit ourselves to two states: an excited state $|k=1\ra$ and the ground state $|k=0\ra$.  The two levels are separated by an energy difference $\hbar \omega_0 = \epsilon( 1) - \epsilon(0)= \eta \pi^2/\beta$ according to (\ref{energy_k2}).  Here we use the notation $|n,k\ra \equiv |n\ra \otimes |k\ra$, where $n$ is the number of photons (not to be confused with $|m\ra$ being the winding states) and  $k \in \mathbb{Z}$ indicates the state of the $\cal{TV}$. Suppose we prepare the system in the $|k=1\ra$ state and tune the oscillating external field to the resonance  frequency $\omega_0$. The transition rate from the $|k=1\ra$ to the $|k=0\ra$ state is  
determined by the corresponding transition matrix element (\ref{transition_element}) inferred previously from  the Euclidean path integral computations (\ref{P_operator}).

First, as  the energy of the k-states grow quadratically with $k$, $\epsilon(k)\sim k^2$, we can neglect highly excited metastable states by considering only the leading contributions to the dipole moment (\ref{transition_element}) due to the transition from  $|k= 1\ra$ to $|k=0\ra$.  To simplify the analysis  and to emphasize the basic features of the system, we also neglect the $|k= -1\ra$  state which is degenerate to the $|k= 1\ra$ state for vanishing external fields. In principle, it  can be easily accounted for. However,  we want to make our formulae as simple as possible, and we ignore this extra state for now. 

If we assume that only a single cavity mode $\omega_0$ exists, which is a good approximation in the case of a high Q resonator, the system can be described by the Jaynes-Cummings Hamiltonian:
\be
&& H=H_0+H_{\rm I} \\
&& H_0=\hbar \omega_0 a^\dagger a + \frac{\hbar \omega_a}{2} \sigma_z , \nonumber\\
&&  H_{\rm I}= \hbar (g a \sigma_+ +g^\ast a^\dagger \sigma_-), \nonumber
\ee
coupling a single harmonic oscillator degree of freedom to our two-level system $|k=0\ra$ and $|k= 1\ra$. Here, $\sigma_{\pm}\equiv\frac{1}{2} (\sigma_x \pm i \sigma_y)$ and $g$ describes the coupling of our two-level system with  the quantized   E\&M field with two transverse polarizations.   Assuming the E\&M field is polarized in the $z$-direction, $g$ reads:
\be
g=-i\sqrt{\frac{\omega}{2 \hbar  V}}   \la k=0| P_z|k=1\ra  .
\ee
One can easily check that on resonance, $\omega_0 = \omega_a$, the interaction Hamiltonian $H_{\rm I}$ commutes with the free Hamiltonian $H_0$, i.e. $[H_0,H_{\rm I}]=0$.  Therefore,  the eigenstates of the full Hamiltonian $H$ can be written as a linear combination of the degenerate eigenstates of $H_0$. The degenerate eigenstates of $H_0$ are $|n,1\ra$ and $|n+1,0\ra$. Within this degenerate subspace, the state of the system at time $t$ can be written $|\Psi(t)\ra = c_{n,1}(t)|n,1\ra +c_{n+1,0}(t) |n+1,0\ra$ and the dressed eigenstates of the full Hamiltonian are $\frac{1}{\sqrt{2}} (|n,1\ra \pm |n+1,0\ra)$. Solving the Schr\"{o}dinger equation yields the time evolution
\be
&& c_{n,1}(t)=c_{n,1}(0) \cos (\Omega_n t) -ic_{n+1,0}(0) \sin(\Omega_n t) \\
&& c_{n+1,0}(t) = c_{n+1,0}(0) \cos (\Omega_n t) -ic_{n,1}(0) \sin(\Omega_n t), \nonumber
\ee
where $\Omega_n=|g|\sqrt{n+1}$.

In particular, if we prepare the $\cal{TV}$ in its excited state $|k=1\ra$ and the initial cavity field with $n$ photons,  i.e. $c_{n,1}(0)=1$ and $c_{n+1,0}(0)=0$, then at a later time $t$ the probability for finding the vacuum in the $|k=1\ra$ state is
\be
P_{k=1}=|\la n,1|\Psi(t)\ra|^2 = \frac{1}{2} (1+\cos 2\Omega_n t).
\ee
The sinusoidal oscillation indicates that energy is constantly exchanged between the $\cal TV$ and the cavity field.
This is of course, the conventional Rabi oscillations with the only difference being that instead of a two-level atomic system, the transitions in our case occur between the metastable and ground states in the $\cal{TV}$, similar to the Brillouin zone classification as discussed at the very beginning of Section \ref{metastable}.

It is particularly interesting to investigate the dynamics of our system ($\cal{TV}$ plus quantum E\&M field) when we  start with an initial cavity field that is a coherent state of photons: 
\be
|\alpha\ra=e^{-{|\alpha|^2}/2} \sum_{n}\frac{\alpha^n}{\sqrt{n!}}|n\ra.
\ee
The time evolution of the $|k=1\ra$ state probability is 
\be
P_{k=1} \approx \frac{1}{2} \left[1+\sum_{n=0}^{\infty}\frac{e^{-|\alpha|^2}|\alpha|^{2n}}{n!}\cos(2\Omega_n t)  \right].
\ee
We conclude this subsection with the following remark. The rate of emission from $\cal{TV}$ due to the non-stationary TCE is very low.
Nevertheless, it is not hopeless to eventually measure this fundamentally new type of energy. The proposal  presented in this section is 
to use resonant cavities for such measurements. A number of historical examples show that such a goal can in principle be achieved\footnote{One can mention a recent  example of the measurement of spontaneous emission
in silicon coupled to a superconducting microwave cavity. The relaxation rate is increased by three orders of magnitude as the spins are tuned to cavity resonance\cite{cavity}.}.

\subsection{Numerical estimates} \label{numerics}

Here, we take the realistic experimental parameters used in \cite{Zhitnitsky:2015fpa,Cao:2015uza} to estimate the dipole transition rates for the magnetic (\ref{correspondence}) and electric (\ref{correspondence1}) systems as well as the lifetime of the excited states. 

For the magnetic system, we use the sample parameters from the experiment on persistent currents \cite{persistent-exp}. The sample in this case was a metallic ring with area $L_1\times L_2=\pi (1.2 \mu {\rm m})^2$ and thickness $L_3=0.1\mu$m at a temperature of $\beta=0.6$cm. The observation of persistent currents implies that coherent Aharonov-Bohm (AB) phases, which is also crucial for the experimental realization of the TCE, can be maintained (see more elaboration on this point in Section VI A in \cite{Cao:2015uza}). Although it was demonstrated in \cite{Zhitnitsky:2015fpa} that for this particular setup, $\tau\gg1$ such that all topological effects are vanishingly small, we here assume that $\tau \sim1$ can somehow be achieved. In this case, the energy separation between the ground ($k=0$) and excited ($k=1$) states is 
\be
\hbar \omega_a \approx 3.2\times10^{-4}~ {\rm eV}.
\ee
The resonant wavelength corresponding to this energy is $3.8$mm, much larger than the dimensions of the system, thereby justifying the dipole approximation in (\ref{correspondence}). The transition rate from the excited to the ground state is then 
\be
R_{1\rightarrow 0} \approx 1.6 \times 10^{-3}~{\rm s}^{-1},
\ee
which corresponds to an excited state lifetime of $6\times10^2$s.

For the electric system, we use the two sets of parameters in \cite{Cao:2015uza}. The first set of parameters is motivated by the  accurate measurement  of the CE using parallel plates \cite{Casimir-exp} (see also 
\cite{Casimir-exp1} where historically the first accurate measurement was performed, but for a different geometry).
The second set of parameters is motivated by the experiments  on persistent currents \cite{persistent-exp} where the correlation of the AB phases is known to be maintained. While the persistent current is a magnetic phenomenon, electromagnetic duality strongly suggests that a similar electric effect should also occur when coherent AB  phases are correlated over macroscopically large distances.\footnote{An interesting impact  of AB phases on tunneling rates have been recently demonstrated in \cite{AB-tunneling}, where photon emission occurs exactly during the tunneling events. The difference from our case is that  the tunneling in our system occurs between distinct  topological vacuum sectors $|m\ra$, while in  Ref.\cite{AB-tunneling} the  charged particle tunnels in the conventional quantum mechanical sense.}
 Therefore, for the second set of parameters we adopt the typical sizes of the magnetic system used above (where persistent currents have been observed) to estimate the topological effects in the electric capacitor configuration. Both sets of parameters can optimize the TCE:
\be
\label{eta1}
\eta^{(I)}=\frac{2 L_1 L_2}{e^2 \beta L_3}=\frac{1.2 \times 1.2 {\rm mm}^2}{2\pi\alpha (180 {\rm mm}) (0.4 {\rm mm})}\approx0.4,
\ee
\be
\label{eta2}
\eta^{(II)}=\frac{2 L_1 L_2}{e^2 \beta L_3}=\frac{2\pi (1.2{\mu m})^2}{4\pi\alpha (0.6 \rm cm) (0.1 \mu m )}\approx0.16.
\ee
The energy separations between the ground ($k=0$) and excited ($k=1$) states are
\be
\hbar \omega_a^{(I)}=4.8\times 10^{-6} {\rm eV}, ~~\hbar \omega_a^{(II)}=5.2\times 10^{-5} {\rm eV}.
\ee
For both sets of parameters, the electric dipole approximation in (\ref{correspondence1}) can be justified by observing that the resonance external electric fields correspond to  wavelengths that are much larger than the sizes of the respective systems:
\be
\lambda^{(I)} \approx 0.3 {\rm m} ~~~~~
\lambda^{(II)} \approx 0.02 {\rm m}.
\ee
 The induced dipole moments   for our  two sets of parameters  can be estimated as \cite{Cao:2015uza}:
 \be
 \label{dipole_numerics}
\langle p^{\rm Mink}_{\rm ind} \rangle^{(I)} &\approx& \frac{eL_3}{2}\eta^{(I)}\sim 0.1 (e \cdot{\rm mm})\nonumber \\
\langle p^{\rm Mink}_{\rm ind} \rangle^{(II)} &\approx& \frac{eL_3}{2}\eta^{(II)}\sim 0.01 (e \cdot{\rm \mu m}).
\ee
These estimates suggest that the effective number of degrees of freedom $n_{\rm eff}$ which coherently generate the dipole moments (\ref{dipole_numerics})
and the corresponding transitions (\ref{transitions_numerics}) can be estimated as $\langle p^{\rm Mink}_{\rm ind} \rangle/(e\cdot 10^{-8}{\rm cm})$, which numerically correspond to $n^{(I)}_{\rm eff}\sim 10^6$ and $n^{(II)}_{\rm eff}\sim 10^2$.

The transition rates (\ref{correspondence1}) for the two sets of parameters are
\be
\label{transitions_numerics}
R_{1\rightarrow 0}^{(I)} \approx 0.21 {\rm s}^{-1}, ~~ R_{1\rightarrow 0}^{(II)} \approx 5.7\cdot 10^{-4} {\rm s}^{-1},
\ee
so the lifetimes of the excited states are 4.7s and $1.7\times 10^{3}$s respectively.

One should emphasize, that in contrast to conventional systems where  a large number of spins are present in a sample, our sets of parameters (\ref{eta1}) and (\ref{eta2}) describe a small, but single macroscopic quantum coherent system. Therefore, a potential detector must be sensitive  to a single photon to observe this new effect
of emission from the $\cal{TV}$.     
 
The rates (\ref{transitions_numerics}), of course, are highly sensitive to all dimensional parameters of the system and the temperature, and show drastic changes when one puts a system into the background of an external field.  In fact, this high sensitivity to external field can   be used to detect the topological vacuum effects as conventional vacuum 
is largely unaffected by any external sources as argued in \cite{Cao:2013na,Cao:2015uza}. Essentially it means that one can scan the system by changing the external   field to search for a resonance response. It also implies that one can, in principle,  manipulate a system in very much the same way as one normally manipulates cold atom systems by tuning the external  field.

\section{Conclusion.   }\label{conclusion}
Our conclusion can be separated into three  related, but still distinct pieces: \\
\ref{basics}. Solid theoretical results  based on the Euclidean path integral computations
in the Maxwell system defined on a compact manifold,  \\
\ref{relation}. Relation to other approaches  where real-time dynamics plays a key role,   and \\
\ref{speculations}. Some  speculations related to strongly coupled QCD realized in nature where fundamentally the same  vacuum effects do occur, and might be the crucial  ingredients   in understanding the observed cosmological vacuum energy. 
In fact, the Maxwell system which is the subject of the present work was originally invented as a theoretical toy model  where some deep theoretical questions can be addressed (and answered) in a simplified setting.

\subsection{Basic results}\label{basics} In this work we discussed a number of very unusual features   exhibited by the Maxwell theory formulated on  a compact manifold $\mathbb{M}$ with nontrivial topological mappings $\pi_1[U(1)]$, termed the  topological vacuum ($\cal{TV}$).  One of the properties which plays an important role in the present studies is the generation of metastable vacuum states, similar to  the classification of Brillouin zones as discussed in Section \ref{metastable}. All these  features originate from the topological portion of the partition function ${\cal{Z}}_{\rm top}$ which is a result of the tunneling events between physically identical but topologically distinct winding states $|n\ra$. The relevant physics    cannot be ascribed to physical propagating photons  with two  transverse polarizations. In other words, all effects discussed in this paper have a ``non-dispersive" nature.

The computations of the present work along with previous calculations of Ref. \cite{Cao:2013na,Zhitnitsky:2013hba,Zhitnitsky:2014dra, Zhitnitsky:2015fpa,Cao:2015uza}  imply  that the extra energy (and entropy),   not associated   with any physical propagating degrees of freedom,  may emerge  in  gauge  systems if some conditions are met. This fundamentally new type  of  energy    emerges as a result of the dynamics of pure gauge configurations and tunneling transitions between physically identical but topologically distinct winding states. The new idea advocated in this work is that this new type of energy can, in principle, be studied  if one places the system in a time-dependent  background, in which  case we expect the topological vacuum  configurations to radiate  conventional propagating photons which can be detected and analyzed according to (\ref{correspondence}) and (\ref{correspondence1}). 
             
As we discussed  in detail in the text, the fundamental technical obstacle   for such an analysis is that the radiation of real physical particles on mass shell is inherently formulated in {\it Minkowski}  space-time with a well-defined Hilbert space of the  asymptotic states. At the same time, the tunneling is described in terms of vacuum fluctuations (``instanton fluxes")  interpolating between the topological   $|n\rangle$  winding sectors and is fundamentally formulated in {\it Euclidean} space-time, see Footnote \ref{QCD} for some comments on this problem. We overcame this technical obstacle  by introducing auxiliary topological fields which, on the one hand, encode the entire information about the tunneling transitions, and on the other hand, can be analytically continued to Minkowski space-time.
Eventually, this approach allowed us to  turn  the problem into conventional Hamiltonian dynamics formulated in Minkowski terms,  as described in Sections \ref{metastable}, \ref{q-transitions}. 
             
The corresponding rate of emission is very low for our system as estimated in Section \ref{numerics}. 
The hope is that microwave cavities may drastically enhance the emission rate such that radiated  photons can be observed.  
Furthermore, putting the system in a background of external electric or magnetic fields, represented by $\theta_{\rm eff}$ in the paper,  one can manipulate  the system  in pretty much the same way as one normally does with cold atom systems by tuning the external field. In practice,  it means that one can scan the system by changing the  external fields to search for a resonance response.
 
\subsection{Relation to other approaches}\label{relation} As emphasized above, we overcame the main technical obstacle in calculating the production of real particles (a real-time process in Minkowski space-time), while dealing with  tunneling processes (formulated in Euclidean space-time) by introducing the auxiliary fields which can be easily continued to Minkowski space-time. This problem is obviously not unique to our work, but  is,  in fact, a common problem when path integrals are performed in Euclidean space-time while  the relevant physical questions are  formulated in  Minkowski terms, see Footnote \ref{QCD}. 

There have been a number of different attempts to attack this problem. The most promising, in our view, is   the approach based on formulating path integrals in Picard-Lefschetz theory. See recent reviews  \cite{Behtash:2015loa,Dunne:2016nmc} and references to the original papers therein. The basic idea there is to formulate real-time path integrals.  The field configurations which describe the tunneling processes live in a complexified field space.  It turns out that the corresponding configurations, being singular, nevertheless produce a finite action for the path integral. In a few simple cases the computations can be explicitly carried out  to reproduce the  known results  in QM systems (see   original computations \cite{Tanizaki:2014xba,Cherman:2014sba,Behtash:2015zha} and reviews  \cite{Behtash:2015loa,Dunne:2016nmc}).  

It is natural to expect that this approach, in principle,  can be generalized to include a time-dependent background field, in which case the complex saddles should be able to  describe tunneling transitions as well as particle   production, precisely the topic of the present work. In other words, we strongly suspect that complex saddles which describe tunneling events in real-time path integrals may also contain information about the production of real particles in a time-dependent background. It  remains to be seen how this information can be recovered from complex saddles.  The answer to this question is not yet known,  as recent  studies \cite{Behtash:2015loa,Dunne:2016nmc}  are mostly focused on analyzing  the properties of the vacuum itself, rather than generalizing this approach to include a time-dependent background to study particle production rates.

\subsection{Speculations}\label{speculations} The unique feature of the system where an extra energy is not related to any physical propagating degrees of freedom was the main  motivation for the proposal   \cite{Zhitnitsky:2013pna,Zhitnitsky:2014aja,Zhitnitsky:2015dia}  that the  vacuum energy of the Universe may have, in fact,  precisely such non-dispersive  nature.\footnote{ This new type  of vacuum energy which can not be expressed in terms of propagating degrees of freedom has in fact been well studied in QCD lattice simulations, see \cite{Zhitnitsky:2013pna} with a large number of references on the original lattice results.} This proposal where an extra energy  cannot be associated with any propagating particles  should be contrasted with the conventional description where an extra vacuum energy in the Universe is always associated with some  ad hoc    propagating degree of freedom.\footnote{There are two instances in the evolution of the Universe when the vacuum energy plays a crucial  role.
The first instance   is identified with  the inflationary epoch  when the Hubble constant $H$ was almost constant, which corresponds to the de Sitter type behavior $a(t)\sim \exp(Ht)$ with exponential growth of the size $a(t)$ of the Universe. The  second instance where the vacuum energy plays a dominant role  corresponds to the present epoch when the vacuum energy is identified with the so-called dark energy $\rho_{DE}$ which constitutes almost $70\%$ of the critical density. In the proposal  \cite{Zhitnitsky:2013pna,Zhitnitsky:2014aja,Zhitnitsky:2015dia}  the vacuum energy density can be estimated as $\rho_{DE}\sim H\Lambda^3_{QCD}\sim (10^{-4}{\rm  eV})^4$, which is amazingly  close to the observed value. }

Essentially, the  proposal   \cite{Zhitnitsky:2013pna,Zhitnitsky:2014aja,Zhitnitsky:2015dia}  identifies the observed vacuum  energy with the topological Casimir type energy, which however is originated not from the dynamics of the physical propagating degrees of freedom, but rather from the dynamics of the topological sectors  that  are always present in gauge systems, and which are highly sensitive to arbitrary large distances. An explicit manifestation of this ``non-dispersive" nature of the vacuum energy in the model considered in the present work is the ``wrong sign" of the kinetic term in the effective Lagrangians describing the dynamics of  the auxiliary no-propagating fields (\ref{Z_auxiliary1}, \ref{Z_auxiliary5}). This ``wrong sign" has exactly the same nature as the conjectured Veneziano ghost introduced in 
QCD to resolve the so-called $U(1)$ problem, see Footnote \ref{ghost} for a few comments on this matter.   Furthermore, the radiation from the vacuum in a time-dependent background (which is  the main subject of this work) is very similar  in all respects to the radiation which might be  responsible for  the  end of inflation in that proposal. The  cosmological ideas of the proposal   \cite{Zhitnitsky:2013pna,Zhitnitsky:2014aja,Zhitnitsky:2015dia}   can hopefully  be tested  in a tabletop experiment (which is the subject of the present paper) where the vacuum energy in a time-dependent background can be transferred to real propagating degrees of freedom as  described in Section \ref{q-transitions}. In cosmology, the corresponding period plays a crucial role and  calls  the reheating epoch which follows inflation with the vacuum energy being the dominant component of the Universe.

To conclude, the main point of the present studies is the claim that the emission of real photons  may occur as a result of   tunneling transitions between topologically distinct but physically identical winding $|n\ra$ sectors, rather than from conventional physical propagating degrees of freedom.

 \section*{Acknowledgements} 
 We are thankful to Charles Cao for the collaboration during the initial stage of the project.    This research was supported in part by the Natural Sciences and Engineering Research Council of Canada. Y.Y. acknowledges a research award from the UBC Work Learn Program.

\appendix
\section{Review of the instanton solutions}\label{appendix_instantons}
In this appendix, we show how the instanton solutions (\ref{toppot4d}) and (\ref{A_top}) are derived. To do so, we first show how they are obtained in the original 2d Maxwell theory (i.e. Schwinger model without fermions) on a toroidal manifold and then extend the results to 4d.

\subsection{2d Maxwell theory}
We follow \cite{Cao:2013na} and ref. therein and solve the 2d Maxwell theory using both the physically transparent Hamiltonian approach and the Euclidean space path integrals with instantons. Their exact agreement validates the use of instantons in this theory.

In the Hamiltonian approach, we define the system on a spatial circile of circumference $L$ at inverse temperature $\beta$. 

We follow the  procedure outlined in \cite{Cao:2013na} and \cite{SM_Hamiltonian} to canonically quantize the 2d Maxwell system. First we fix the gauge: 
\be
\label{gauge}
A_0=0 ~~~~\partial_1 A_1=0.
\ee  
Hence, $A_0$ is not a dynamical variable. On the other hand, $E= \dot{A_1}(t)$.
We impose conventional periodic boundary conditions:
\be
\label{bc}
A_1 (t, ~x=-\frac{L}{2})= A_1 (t, ~x=\frac{L}{2}).
\ee 

The theory is defined by the following Hamiltonian density and commutation relations:
\be
{\cal H}=\frac{1}{2} E^2
\ee
\be
\label{commutator}
[A_1(x),E(y)]=i\delta(x-y).
\ee
We also need to impose Gauss law on the set of physical states $|{\rm phys} \ra$:
\be
\label{gauss}
\partial_1 E |{\rm phys} \ra=0,
\ee 
which is only satisfied by the $x$-independent zero mode. As it is known, there is 
a class of admissible gauge transformations, the so-called large gauge transformations
\be
\label{gauge_transform}
A_1\rightarrow A_1+\frac{d\alpha (x)}{dx}, ~~~ \alpha=\frac{2\pi n x}{eL}, ~~~ n=\pm 1, \pm 2 ... ~  . ~~~~
\ee
This gauge function is compatible with periodic boundary conditions (\ref{bc}) because $ {d\alpha (x)}/{dx}=$ const,  and the periodicity  (\ref{bc})  is not violated. 
This implies the following gauge equivalence relation
\be
\label{compactification}
A_1 \sim A_1 + \frac{2\pi }{e L}n.
\ee
Hence, we conclude that $A_1$ is not independent on the entire interval $(-\infty,\infty)$ and instead lives on a circle of circumference $2\pi/eL$. 

By expanding $A_1$ and $E$ in their Fourier modes, we can map the current problem onto the particle on a ring problem in quantum mechanics. The conjugate momentum operator and the Hamiltonian read
\be
\label{operator}
&& E=-\frac{i}{L} \frac{d}{dA},\\
&& H=\frac{L}{2} (-\frac{i}{L} \frac{d}{dA})^2.
\ee
$H$ acting on the energy eigenstates $\exp{(ienLA)}$ yields the eigenvalues $\epsilon_n=\frac{1}{2}n^2 e^2 L$. The partition function at inverse temperature $\beta$ is therefore given by the canonical ensemble
\be
\label{Z_Hamiltonian_approach}
{\cal Z}={\rm tr}~e^{-\beta H}=\sum_n e^{-\beta \epsilon_n}=\sum_n \exp{(- \frac{\beta e^2 L}{2}n^2 )}, ~~
\ee
where we have taken the fundamental theta term, $\theta=0$ to simply formulae and notations. 

In the path integral approach (\cite{Cao:2013na} and \cite{SW,Azakov}), we solve the same problem in Euclidean space-time with metric $(1,1)$. Time and space form a two-torus with size  $\beta \times L$. In the context of this problem, the topology of the system is equivalently taken into account by imposing periodic boundary conditions up to a large gauge transformation and using the so-called instanton solutions.

The Maxwell equations with the appropriate boundary conditions, for instance
\be
\label{2d_BC}
A_\mu(x_0,x_1+L)=A_\mu(x_0,x_1)+\partial_\mu \frac{2\pi k}{\beta} x_0,
\ee
yield solutions of the form
\be
\label{gauge_decomposition}
A^{(k)}_{\mu} = A^{(0)}_{\mu} + C^{ (k)}_{\mu},
\ee
where $A^{(0)}_{\mu}$ is the exactly periodic quantum field, and $C^{ (k)}_{\mu}$ is the classical instanton solution. 

In Lorenz gauge, the instanton solution can be written 
\be
\label{2d_instanton}
A^{{\rm top}}_\mu=C^{ (k)}_{\mu}=(-\frac{\pi k}{eV}x_1, ~\frac{\pi k}{eV}x_0)
\ee
where $V=\beta L$ is the volume of the Euclidean space-time. These instanton configurations classified by integers $k$  describe tunneling between different vacuum winding states, say $|m\ra$ and $|m'\ra$ with $k=m'-m$ (cf. Eq. (\ref{large_gauge_transform})). They also give rise to a topological electric field
\be
\label{2d_E}
E_{\rm top}=\partial_0 A^{{\rm top}}_1 - \partial_1 A^{{\rm top}}_0=\frac{2 \pi k}{eV}.
\ee 
It is worth mentioning that the topological electric field (\ref{2d_E}) should not be confused with the familiar physical electric field in Minkowski space-time, which is the eigenvalue of the $E$ operator  ({\ref{operator}).    Rather, it is an effective electric field in the unphysical Euclidean space-time and is better thought of as some complex configuration that saturates the Euclidean path integral and that describes tunneling transitions between distinct topological sectors.  In particular, the dependence of these fields on the coupling constant $e$ is drastically different:  the topological $E_{\rm top} $ configuration describing the tunneling amplitude   is proportional to $ e^{-1}$, while physical electric field being the eigenvalue of  ({\ref{operator}) is proportional to $e$. 

The partition function can be obtained by doing the following path integral and explicitly summing over topologies
\be
\label{path_integral}
{\cal Z}=\sum_{k \in \mathbb{Z}} \int {\cal D}A_\mu^{(k)}~ e^{ \int\dd^2 x( -\frac{1}{2}E^2 )}.
\ee
Here, $E$ includes both the quantum fluctuations and the topological field (\ref{2d_E}).

Omitting the computational details, the partition function is
\be
\label{Z_top_2D}
{\cal Z}={\cal Z}_{\rm quant} \times {\cal Z}_{\rm top}=\sqrt{\frac{2 \pi}{e^2 V}} \sum_{k \in \mathbb{Z}} e^{-\frac{2\pi^2 k^2}{e^2 V} }.
\ee

Although this partition function (\ref{Z_top_2D}) looks different from the one obtained earlier in the Hamiltonian approach (\ref{Z_Hamiltonian_approach}), they are in fact dual expressions of each other related by the Poisson summation formula. Thus, although it is not straightforward how one can directly relate the boundary conditions in the Hamiltonian approach (\ref{bc}) (i.e., strictly periodic) to those in the path integral approach (\ref{2d_BC}) (i.e., periodic up to a large gauge transformation, giving rise to the instantons), their agreement in the end validates the use of instantons.   In fact, the relation between these two approaches is quite complicated, see detailed analysis in \cite{Azakov}, and  also related  discussions in    \cite{ChenLee}. 

Our computational framework in the main body of this paper is entirely based on Euclidean path integrals. 
Therefore, in  this framework  we impose the boundary conditions up to large gauge transformations, similar to the above discussions. The corresponding fields, such as (\ref{2d_E}),  should be interpreted as the  field configurations (describing the tunneling processes between the  topological sectors)   saturating the path integral, not to be confused with real fields representing the eigenvalues of the system, as we already mentioned after Eq. (\ref{2d_E}).

\subsection{4d Maxwell theory}
If we consider the Maxwell theory in 4d space-time, the topologies of the space-time becomes substantially more complicated. The same periodic boundary conditions up to a gauge potentially yields six different instanton solutions, corresponding to the 6 hypersurfaces in 4d. However, if require two of the dimensions of space-time to be much greater than the other two, we essentially dimensionally reduce the problem to the previous 2d problem.  Again there are six ways this can be done, and the electric and magnetic cases discussed in Sect. \ref{magnetic} and \ref{electric} are precisely two of them.

$\bf{Case~1:}$ $\beta,L_3\ll L_1,L_2$. The dominant instanton and the corresponding bouundary conditions are a straightforward generalization of (\ref{2d_BC}, \ref{2d_instanton}):
\be
\label{A_top_B}
A^{\mu}_{\rm top} = \left(0 ,~ -\frac{\pi k}{e L_{1} L_{2}} x_2 ,~ \frac{\pi k}{e L_{1} L_{2}} x_1 ,~ 0 \right).
\ee  
This instanton configuration gives rise to a uniform ``topological magnetic field'' in the $z$ direction:
\be
\label{B_top}
\vec{B}_{\rm top} &=& \vec{\nabla} \times \vec{A}_{\rm top} = \left(0 ,~ 0,~ \frac{2 \pi k}{e L_{1} L_{2}} \right),\\
\Phi&=&e\int \dd x_1\dd x_2  {B}_{\rm top}^z={2\pi}k. \nonumber
\ee

$\bf{Case~ 2:}$ $\beta,L_3\gg L_1,L_2$. The instanton that contributes the most will be
\be
\label{A_top_E}
A^3_{\rm top}(\beta)&=&A^3_{\rm top}(0)+\frac{2\pi k}{eL_3}, \\  \nonumber
A^{\mu}_{\rm top} (t) &=& \left(0 ,~ 0,~ 0, ~ \frac{2\pi k}{e L_{3} \beta } t  \right),
\ee  
which produces a uniform ``topological electric field'' in the $z$ direction:
\be
\label{E_top_1}
\vec{E}_{\rm top} &=&\dot{\vec{A}}_{\rm top} = \left(0 ,~ 0,~ \frac{2 \pi k}{e L_{3} \beta} \right),\\
\Phi&=&e\int \dd t \dd x_3  {E}_{\rm top}^z={2\pi}k. \nonumber
\ee

Certainly, the instanton solution given in (\ref{A_top_E}) still exists in a system with the first set of dimensional reduction conditions $\beta,L_3\ll L_1,L_2$, but the resulting action $S_2\sim \int \dd^4 x (1/\beta L_3)^2$ will be much large than that of the first type of instanton, $S_1 \sim \int \dd^4 x(1/L_1 L_2)^2$. 

As in the 2d case, it is far from obvious how one can formulate the boundary conditions in real 4d Minkowski space-time as in the Hamiltonian approach and then explicitly derive the above boundary conditions for the Euclidean instantons. These instantons should be treated as auxiliary field configurations saturating the path integral, and such an interpretation is further supported by our studies  in the present work where the configurations saturating the path integral are in fact complex-valued fields,  which obviously cannot be confused with real physical configurations. However, the exact analogy between the 2d and 4d cases achieved via dimensional reduction\footnote{For example, we explicitly see that the topological portion of the partition function for the 4d electric case (\ref{Z_M}) reduces to the 2d partition function (\ref{Z_top_2D}) in the limit $L_1, L_2 \rightarrow 0$ accompanied by a proper rescaling of the coupling constant $e$.} strongly  suggests that similar to the simple boundary conditions in the 2d Hamiltonian method (\ref{bc}), all we need for real experiments in 4d Minkowski space-time is periodic boundary conditions in the relevant directions (without large gauge transformations). 

In short, we require periodicity up to large gauge transformations to perform mathematical derivations in Euclidean space-time, whereas simple periodic boundary conditions are needed in Minkowksi space-time, both for Hamiltonian solutions and for experiments.

 It is quite possible that formulating the path integral in Picard-Lefschetz theory using Minkowski space-time from the start, as mentioned in Section \ref{relation}, may give a precise answer to the relation between these two descriptions.   However,  the corresponding computational framework  is not presently known, and  yet to be developed. 

\section{On  possible design of the quantum LC circuits in real Minkowski space-time}\label{design}
We  make a few comments here on the possible design of a system satisfying the periodic boundary conditions that represent the key element for generating $\cal{Z}_{\rm top}$ from nontrivial $\pi_1[U(1)]$. As we emphasized  in Appendix \ref{appendix_instantons}, the construction in Minkowski space-time requires  simple  periodic boundary conditions.
It is only our mathematical construction  of the Euclidean path integral that requires 
more complicated boundary conditions (periodic up to large gauge transformations), which produce  gauge images of the original 
interval where the gauge  field is defined. In the path integral approach, the summation of an infinite number of gauge images is harmless as any expectation value is always  computed by normalizing to the same partition function which also includes the same  infinite sum. 
  
The subject of the present Appendix is a possible design in Minkowski space-time. Therefore, we do not discuss Euclidean instantons, nor configurations that saturate the Euclidean path integral. Instead, we focus on the physics  in Minkowski space with exactly periodic boundary conditions (\ref{bc}) and without summation over the gauge images.   Experience with the 2d model reviewed in Appendix \ref{appendix_instantons} shows  that in Minkowski space-time, such boundary conditions  do generate the topological physics studied in the present work.
   
{\bf Case 1.} The simplest way to realize the periodic boundary conditions in magnetic systems is to make a cylinder as discussed in great detail 
in \cite{Zhitnitsky:2015fpa}. One can explicitly see that the fluctuating magnetic fluxes can be formulated in terms of boundary currents
flowing along the cylinder. In many respects the physics is very similar to (but still distinct from, see  \cite{Zhitnitsky:2015fpa} for details) the   persistent currents observed in a number of materials including metals, insulators, and semiconductors.  In particular, the corresponding instanton fluxes would fluctuate even without external magnetic field,
in contrast to conventional persistent currents which occur exclusively due to the external magnetic field.    
The key requirement is, of course,  that Aharonov-Bohm coherence be maintained in the entire system. 

{\bf Case 2.} The electric systems can be realized with a small capacitor consisting of two parallel plates with plate area $L_1 \times L_2 $ and separation $L_3$, such as the mentioned in Sections \ref{electric} and \ref{numerics}. We connect the two plates with a superconducting wire. In the  wire, the electromagnetic fields vanish, and the gauge fields $A_\mu$ must be a constant or its gauge transform. Therefore,  the wire essentially identifies the two plates and enforces $A_\mu$ to be the same on the plates,  giving conventional boundary conditions similar to (\ref{bc}):
\be
\label{bc_4d}
A_{\mu} (t, ~z=0)= A_{\mu} (t, ~z=L_3). 
\ee 
 The large gauge transformations along $z$ 
\be
\label{gauge_transform_4d}
A_3&\rightarrow& A_3+\frac{d\alpha (z)}{dz}, ~~~~~~~~ \psi\rightarrow e^{i\alpha(z)}\psi \nonumber\\
  \alpha&=&\frac{2\pi n z}{eL_3}, ~~~~~~~~ n=\pm 1, \pm 2 ... \nonumber
\ee
obviously respect  the boundary conditions (\ref{bc_4d}) because $ {d\alpha (z)}/{dz}=$ const.
Similar to the 2d analysis, we conclude that $A_3$ field  lives on a circle of circumference $2\pi/eL_3$. Thus, the 2d system represents a dimensionally reduced version of the current 4d electric system, as the topological portion of the partition function for the 4d electric case (\ref{Z_M}) reduces exactly to the 2d partition function (\ref{Z_top_2D}) in the limit $L_1, L_2 \rightarrow 0$ accompanied by a proper rescaling of the coupling constant $e$. Hence, the similar $\mathbb{S}^1$ topological physics will be generated in the 4d system, which gives rise to the physically observable phenomena discussed in this paper.


\begin{thebibliography}{99}
   	
   	\bibitem{Cao:2013na} 
   	C.~Cao, M.~van Caspel and A.~R.~Zhitnitsky,
   	Phys.\ Rev.\ D {\bf 87}, 105012 (2013)
   	[arXiv:1301.1706 [hep-th]].
   	
   	\bibitem{Zhitnitsky:2013hba} 
   	A.~R.~Zhitnitsky,
   	Phys.\  Rev.\  D {\bf 88}, { 105029} (2013)
   	[arXiv:1308.1960 [hep-th]].
   	
   	\bibitem{Zhitnitsky:2014dra} 
   	A.~Zhitnitsky,
   	Phys.\ Rev.\ D {\bf 90},  105007 (2014)
   	[arXiv:1407.3804 [hep-th]].
   	
   	\bibitem{Zhitnitsky:2015fpa} 
   	A.~R.~Zhitnitsky,
   	Phys.\ Rev.\ D {\bf 91}, 105027 (2015)
   	[arXiv:1501.07603 [hep-th]].

\bibitem{Cao:2015uza} 
  C.~Cao, Y.~Yao and A.~R.~Zhitnitsky,
  Phys.\ Rev.\ D {\bf 93}, 065049 (2016)
  [arXiv:1512.00470 [hep-th]].
  
 \bibitem{Casimir}
   H. B. G. Casimir, Kon. Ned. Akad. Wetensch. Proc. {\bf 51}, 793 (1948).
  
\bibitem{Cho:2010rk} 
  G.~Y.~Cho and J.~E.~Moore,
  Annals Phys.\  {\bf 326}, 1515 (2011)
  [arXiv:1011.3485 [cond-mat.str-el]].

\bibitem{Wen:2012hm} 
  X.~-G.~Wen,
  arXiv:1210.1281 [cond-mat.str-el].
  
\bibitem{Sachdev:2012dq} 
  S.~Sachdev,
  arXiv:1203.4565 [hep-th].
  
   
\bibitem{Cortijo:2011aa} 
  A.~Cortijo, F.~Guinea and M.~A.~H.~Vozmediano,
  J.\ Phys.\ A {\bf 45}, 383001 (2012)
  [arXiv:1112.2054 [cond-mat.mes-hall]].

\bibitem{Volovik:2011kg} 
  G.~E.~Volovik,
  Lecture Notes in Physics, {\bf 870}, 343 (2013)
  [arXiv:1111.4627 [hep-ph]].
  
 
  
  \bibitem{SW}
 I.~Sachs and A.~Wipf,
 Helv.\ Phys.\ Acta {\bf 65}, 652 (1992);\\
 I.~Sachs and A.~Wipf,
 Annals Phys.\  {\bf 249}, 380 (1996)
 [arXiv:hep-th/9508142];\\
 S.~Azakov, H.~Joos and A.~Wipf,
 Phys.\ Lett.\ B {\bf 479}, 245 (2000)
 [hep-th/0002197].
 \bibitem{Azakov}
 S.~Azakov,
 Int.\ J.\ Mod.\ Phys.\ A {\bf 21}, 6593 (2006)
 [hep-th/0511116].
 
  


  \bibitem{DCE}
  G. T. Moore, J. Math. Phys. {\bf 11}, 2679 (1970); \\
  S. A. Fulling and P. C. W. Davies, Proc. R. Soc. Lond. {\bf A 348}, 393 (1976); \\
  P. C. Davies and S. A. Fulling, Proc. Roy. Soc. Lond. {\bf A 356}, 237 (1977).
  
  \bibitem{DCE-review}
  G. Barton and C. Eberlein, Ann. Phys. {\bf 227}, 222 (1993);\\
  M. Kardar et al., Rev. Mod. Phys. {\bf 71}, 1233 (1999);\\
       V. V. Dodonov, pp. 309 in Modern Nonlinear Optics, Part 3, ed. M. W. Evans, Adv. Chem. Phys. Series, Vol. {\bf 119} (Wiley, New York, 2001). 
  
  
  \bibitem{DCE-exp}
  C. M. Wilson, G. Johansson, A. Pourkabirian, M. Simoen, J. R. Johansson, T. Duty, F. Nori
and P. Delsing, Nature, {\bf 479}, 376 (2011);\\
  P.Lahteenmaki, G.S.Paraoanu, J.Hassel, and P.J.Hakonen,
Proc. Natl. Acad. Sci. U.S.A. {\bf 110}, 4234 (2013).

 

\bibitem{BF}
T.H.Hansson, V. Oganesyan, and S.L.Sondhi, Annals of Physics {\bf 313}, 497 (2004)
arXiv:cond-mat/0404327


\bibitem{Zhitnitsky:2013hs} 
  A.~R.~Zhitnitsky,
  Annals Phys.\  {\bf 336}, 462 (2013)
  [arXiv:1301.7072 [hep-ph]].
  
      \bibitem{ChenLee}
  K.T.~Chen and P.A.~Lee, 
    Phys.\ Rev.\ B {\bf 83}, 125119 (2011)
  [arXiv:1012.2084].

\bibitem{Shifman}
M.A. Shifman, \textit{Advanced Topics in Quantum Field Theory: A Lecture Course}, Cambridge and New York, Cambridge University Press, 2012.

\bibitem{Witten:1980sp} 
  E.~Witten,
  Annals Phys.\  {\bf 128}, 363 (1980).

\bibitem{wittenflux}
  E.~ Witten,  
  Phys. \ Rev.  \ Lett. {\bf  81}
  2862 (1998) [arXiv:hep-th/9807109].

 
\bibitem{Bhoonah:2014gpa} 
  A.~Bhoonah, E.~Thomas and A.~R.~Zhitnitsky,
  Nucl.\ Phys.\ B {\bf 890}, 30 (2014)
  [arXiv:1407.5121 [hep-ph]].
   
    \bibitem{cavity}
  A. Bienfait et al, Nature {\bf 531}, 74-77, (2016)
 
  \bibitem{persistent-exp}
  V. Chandrasekhar, R. A. Webb, M. J. Brady, M. B. Ketchen, W. J. Gallagher, and A. Kleinsasser, 
  Phys. \ Rev.  \ Lett.,  {\bf 67},   3578 (1991).
 
  
  \bibitem{Casimir-exp} G. Bressi, G. Carugno, R. Onofrio, and G. Ruoso  Phys. Rev. Lett. {\bf 88}, 041804 (2002).
  
  \bibitem{Casimir-exp1} S.K.~Lamoreaux, Phys. Rev. Lett. {\bf 78}, 5 (1997)
  
  \bibitem{AB-tunneling}
  Atsushi ~Noguchi, Yutaka ~Shikano, Kenji ~Toyoda, and Shinji ~Urabe,
  Nature Communications {\bf 5}, 3868 (2014)
   [arXiv:1405.5052 [quant-ph]].

   
\bibitem{Behtash:2015loa} 
  A.~Behtash, G.~V.~Dunne, T.~Schaefer, T.~Sulejmanpasic and M.~Unsal,
  arXiv:1510.03435 [hep-th].
  
\bibitem{Dunne:2016nmc} 
  G.~V.~Dunne and M.~Unsal,
  arXiv:1601.03414 [hep-th].
  
\bibitem{Tanizaki:2014xba} 
  Y.~Tanizaki and T.~Koike,
  Annals Phys.\  {\bf 351}, 250 (2014)
  [arXiv:1406.2386 [math-ph]].
  
\bibitem{Cherman:2014sba} 
  A.~Cherman and M.~Unsal,
  arXiv:1408.0012 [hep-th].
  
\bibitem{Behtash:2015zha} 
  A.~Behtash, G.~V.~Dunne, T.~Schäfer, T.~Sulejmanpasic and M.~Ünsal,
  Phys.\ Rev.\ Lett.\  {\bf 116}, no. 1, 011601 (2016)
  [arXiv:1510.00978 [hep-th]].


\bibitem{Zhitnitsky:2013pna} 
  A.~R.~Zhitnitsky,
  Phys.\ Rev.\ D {\bf 89}, 063529 (2014)
  [arXiv:1310.2258 [hep-th]].
  
\bibitem{Zhitnitsky:2014aja} 
  A.~R.~Zhitnitsky,
  Phys.\ Rev.\ D {\bf 90}, 043504 (2014)
  [arXiv:1404.5965 [hep-ph]].


\bibitem{Zhitnitsky:2015dia} 
  A.~R.~Zhitnitsky,
  Phys.\ Rev.\ D {\bf 92}, no. 4, 043512 (2015)
  [arXiv:1505.05151 [hep-ph]].
  
\bibitem{SM_Hamiltonian}
N.S. Manton, Ann. Phys. {\bf 159}, 220 (1985);\\
R. Linares, L.F. Urrutia, and J.D. Vergara, (2000). [arXiv:hep-th/0010114 [hep-th]].  




   \end{thebibliography}
\end{document}